\documentclass[11pt]{article}
\usepackage{jheppub}
\pdfoutput=1

\usepackage{slashed}
\usepackage{array,multirow}
\usepackage{graphicx,graphics}
\usepackage{amsfonts,amsmath,amssymb, bbold,bbm,subcaption}
\usepackage{color,xcolor}
\usepackage{tikz} 
\usepackage{enumitem}
\usepackage{cleveref}
\usepackage{hyperref}

\newcommand{\be}{\begin{equation}}
\newcommand{\ee}{\end{equation}}
\renewcommand\({\left(}
\renewcommand\){\right)}
\renewcommand\[{\left[}
\renewcommand\]{\right]}
\newcommand{\dd}{{\rm d}}

\newcommand{\e}{{\rm e}}
\newcommand\vp{\varphi}
\newcommand\eps{\epsilon}

\newcommand\mpl{m_{\rm pl}}

\let\vec\mathbf
\let\vecS\boldsymbol

\def\L{\mathcal{L}}
\def\O{\mathcal{O}}
\def\M{\mathcal{M}}

\def\A{\mathcal{A}}
\def\P{\mathcal{P}}
\def\F{\mathcal{F}}

\def\nn{\nonumber}
\def\Tr{{\rm Tr}}

\preprint{ Nikhef-2022-025}

\title{SIMPly add a dark photon}

\abstract{
  Pions of a dark sector gauge group can be strongly interacting massive particle (SIMP) dark matter, produced by the freeze-out of $3 \to 2$ interactions, with naturally large self-interactions.  We study if adding a dark photon to the set-up can do it all: i) maintain thermalization with the visible sector, ii) resonantly enhance the $3\to2$ interactions, thus allowing for a perturbative pion description, and iii) provide a velocity dependent self-interaction that can affect small scale structure formation.  For $N_f=3$ this minimal setup is marginally excluded, as the required kinetic mixing is too small to maintain thermal equilibrium with the SM. Adding an extra dark quark opens up parameter space, and -- perhaps somewhat surprisingly -- we find that all bounds can be satisfied for dark pion masses $m_\pi \sim 250-600\,$MeV. Dropping the small scale structure requirement iii), a viable setup is reproduced for dark charges of $\alpha_d = 0.01-1$ and a dark pion mass $m_\pi \geq 30$ MeV. Late time annihilations are non-negligible making the SIMP dark pion a bit WIMPy.

}

\author[a,b]{Pieter Braat}
\author[a,c]{and Marieke Postma}
\emailAdd{pbraat@nikhef.nl, mpostma@nikhef.nl}
\affiliation[a]{Nikhef, Theory Group, Science Park 105, 1098 XG Amsterdam, The Netherlands}
\affiliation[b]{Institute of Physics, University of Amsterdam, Science Park 904, 1098 XH Amsterdam, The Netherlands}
\affiliation[c]{Institute for Mathematics, Astrophysics and Particle Physics, Radboud University
Nijmegen, Heyendaalseweg 135, Nijmegen, the Netherlands}

\begin{document}

\maketitle

\section{Introduction}

Despite ongoing experimental and theoretical efforts, the nature of DM remains elusive.  An attractive possibility is that DM is a thermal relic, and its abundance is determined by freeze-out from the thermal plasma. Although most attention has been on weakly interacting particles (WIMPs), their
parameter space is increasingly constrained \cite{LZ:2022ufs,Aprile_2021,PandaX:2022osq}, and other explanations have come to prominence. One such alternative is strongly interactive massive particles (SIMPs) \cite{Hochberg:2014kqa,Hochberg:2015vrg,Lee:2015gsa}. In the SIMP scenario, dark matter freeze-out occurs in a dark sector via $3\rightarrow2$ interactions, which are typically stronger than in the WIMP scenario. As a result, SIMPs have a lower mass (typically MeV-GeV scale), to which direct detection experiments are less sensitive \cite{Hochberg:2014dra}. To avoid overheating the dark sector during freeze-out, a portal coupling maintains thermal equilibrium with the visible Standard Model sector.

Cosmological observations question the collisionless dark matter (CDM) paradigm. For example, observations of dark matter halo density profiles do not match the expected NFW-profile \cite{Navarro:1996gj,Navarro:1995iw} of CDM \cite{Moore:1994yx,Flores:1994gz,Walker_2011,deBlok:2001hbg,deBlok:2002vgq,Simon:2004sr}. This discrepancey is known as the cusp vs. core problem, see e.g. \cite{DelPopolo:2021bom} for a recent review. Although the inclusion of baryonic effects in the numerical simulations may resolve the discrepancy \cite{Navarro:1996bv,Gelato:1998hb, Mashchenko:2007jp, Governato:2009bg, Governato:2012fa}, it is also possible that the resolution lies in the dark matter properties.  Indeed, dark matter with strong self-interactions naturally alleviate the problem by 
transferring heat from the inner to the outer parts of the halo, thus smoothening the density profile \cite{Spergel:1999mh,Kaplinghat:2015aga}. The required interactions are scale dependent --  a factor 10 difference between galaxies and galaxy clusters -- pointing to a velocity-dependent self-interaction \cite{Kaplinghat:2015aga}.

The archetypical SIMPs are the pseudo Nambu-Goldstone bosons -- the dark  pions -- of a condensed dark Yang-Mills theory, with the Wess-Zumino-Witten term providing the five-point interactions \cite{Wess:1971yu,Witten:1983tw,Witten:1983tx}. The dark pions have naturally large self-interactions, and may address the small scale problems of CDM as well \cite{Cline:2013zca}, except that the interactions are not velocity dependent. Moreover, satisfying both the relic density and the self-interaction constraints (or more conservatively, the upper bound on the self-interactions from the Bullet cluster observations), is only possible for non-perturbatively large pion couplings, invalidating the chiral perturbation theory approach \cite{Hochberg:2014kqa}.  Both of these issues can be overcome with extra vector bosons in the model \cite{Choi:2018iit}, and in this paper we will consider adding a massive dark photon. For finetuned dark photon mass, almost twice the dark pion mass, the photon mediated self-interactions can be on resonance, thus giving rise to a velocity-dependent effect \cite{Chu:2018fzy}.  Moreover, the WZW-interactions may likewise be resonantly enhanced, increasing the freeze-out interactions \cite{Maity:2019vbo, Dey:2018yjt}. 

In addition, the dark photon can maintain equilibrium with the SM through kinetic mixing \cite{Okun:1982xi,Holdom:1985ag} with the SM photon \cite{Lee:2015gsa,Hochberg:2015vrg}.  In this case dark pion annihilations to SM final states should be included in the relic density calculations as well.  The annihilation rate grows at late times, as the dark pions lose kinetic energy and the annihilation cross section gets more and more resonantly enhanced.  As a result, even after freeze-out of the  (resonantly enhanced) WZW interactions, the annihilations can still be important and affect the final relic density. Late time annihilations in photons and electrons are bounded by nucleosynthesis and cosmic microwave background observations.

In this paper we will study if one particle -- the dark photon -- can do it all, solve the small scale structure problems of CDM, enhance the freeze-out interactions such that the relic density is obtained for non-perturbative pion couplings, and maintain thermal equilibrium with the SM -- this is dubbed the resonant self-interacting dark matter (RSIDM) scenario.  We also include collider bounds on dark photon kinetic mixing
, as well as cosmological constraints from nucleosynthesis and the cosmic microwave background.  We find that RSIDM can affect small scale structure formation while maintaining thermal equilibrium with the SM for the non-minimal setup with four dark quarks. The minimal setup with three quarks is only marginally excluded by the thermalisation requirement. In both setups more precise calculations may be needed to make definite statements. We will also study how parameter space opens up if the requirement that dark matter affects small scale structure formation is dropped.


This paper is structured as follows.  \Cref{sec:lagrangian} introduces the dark sector, with the dark pions and dark photon.  This is followed by a discussion of the dark matter self-interactions and Bullet cluster bound in \cref{sec:self-int}; analytical estimates for the freeze-out temperature and final relic density, including both WZW interactions and annihilations, in \cref{sec:freeze-out};  and the constraints on the kinetic mixing parameter in \cref{sec:mixing}.  In \cref{sec:scenarios} we then discuss the parameter space for which the the relic density can be obtained in a perturbative set-up, the self-interactions can be resonantly enhanced to address the small scale structure problems of CDM, and the dark photon can keep the dark and visible sector in thermal equilibrium during freeze-out. In addition to the analytical estimates we will also provide numerical results. We end with concluding remarks in \cref{sec:conclusion}.  For completeness, we have also added the computation of the various (thermally averaged) cross sections in the appendix. Our results for  the WZW and pion self-interactions agree with the literature; new is the photon mediated resonant contributions to the various cross section. 

\section{Lagrangian}
\label{sec:lagrangian}

Strongly interacting massive particles (SIMPs) freeze out via $3 \to 2$ dark matter interactions \cite{Hochberg:2014dra,Bernal:2015bla}.  The large required number changing interactions can naturally be obtained in a dark sector with a non-abelian symmetry, with dark pions playing the role of the DM \cite{Hochberg:2014kqa}. In this paper we study the phenomenology of this set-up if we add a dark photon \cite{Lee:2015gsa,Hochberg:2015vrg}. The dark photon can provide a portal between the dark and SM sectors, and -- for tuned masses -- can resonantly enhance the dark pion interactions. 

We thus consider a dark sector with an $SU(N_c)\times U(1)$ gauge symmetry, with $N_f$ dark quarks in the fundamental representation of the gauge group. The quark mass matrix is assumed diagonal $M = m_q \mathbb{1}$, allowing for a Wess-Zumino-Witten (WZW) term in the action for $N_c \geq3$ colors \cite{Wess:1971yu,Witten:1983tw,Witten:1983tx}.  A dimension-four kinetic mixing operator connects the dark $U(1)$ group with the SM hypercharge \cite{Okun:1982xi,Holdom:1985ag}, and provides a portal between the dark and visible sectors.

At a scale $\Lambda$ the non-abelian gauge group condenses, and the approximate flavor symmetry of the light left- and right-handed quarks is broken down to the diagonal subgroup. 
The $(N_f^2-1)$ dark pions are the pseudo-Goldstone bosons of the this symmetry breaking. They can be naturally lighter than the scale $\Lambda$, which sets the mass of the baryons in the theory. Depending on their couplings to other sectors, including the SM, the dark pions can be stable on cosmological timescales and thus are a good dark matter candidate.

At energies below the condensation scale the effective action is
\be
S =\int \dd^4 x \, \[ \frac{f_\pi^2}{4} \Tr (D_\mu U)^\dagger D^\mu U +\frac{\zeta f_\pi^3}{2}
  \Tr (M U +{\rm h.c.}) -\frac14 V_{\mu\nu}^2 - \frac12 m_V V_\mu^2 - \eps V_\mu J^\mu_{\rm SM} \] + \Gamma_{\rm WZW}
\label{L}
\ee
with $\zeta ={\cal O}(1)$.  The first two terms are the leading order operators of the chiral effective Lagrangian describing the dark pion dynamics.  Here $U = \e^{2i \vecS \pi/f_\pi}$, $\vecS \pi = \pi^a T^a$ with $T^a$ generators 
of $SU(N_f)$, and $f_\pi \sim \sqrt{N_c} \Lambda/(4\pi)$ the pion decay constant.
The covariant derivative is $D_\mu U = \partial_\mu U + ig_d [Q,U] V_\mu$, with $V_\mu$ the dark photon field and $g_d$ the dark U(1) gauge coupling.  We choose the charge matrix  \cite{Bar:2001qk,Lee:2015gsa}  %
\be
Q ={\rm Diag}(1,-1,1,-1,...),
\label{Q}
\ee
with $N_f$ entries. With this charge assignment $\Tr(Q^2 T^a)=0$ and the mixed anomaly vanishes, avoiding decay of the neutral dark pion into to two (dark) photons \cite{Bardeen:1969md,Ioffe:2007eg}. 

The next two terms are the dark photon kinetic term, where we defined the gauge field strength $V_{\mu\nu} =\partial_\mu V_\nu-\partial_\nu V_\mu$, and the St\"uckelberg mass term for the dark photon. To avoid pion decay the dark photon mass should exceed twice the pion mass;  in the limit that the photon mass is close  to that  threshold, the photon-pion interactions can be resonantly enhanced.  We parameterize\footnote{Here $m_\pi$ is the mass of the charged -- with unit charge -- dark pions, which interact with the dark photon. The mass of the charged dark pions receive loop corrections, and is slightly larger than the mass of the neutral pions. }
\be
m_V = m_\pi(2+ \delta m)> 2m_\pi.
\label{mV}
\ee
The last term in \cref{L} between the square brackes couples the dark photon to the SM vector current $J^\mu_{\rm SM} =\sum q_f \bar f \gamma^\mu f$, and $q_f$ the electric charge of the SM fermion $f$. This term arises as a consequence of kinetic mixing between the dark $U(1)$ group and SM hypercharge; after redefining the fields to make the kinetic terms canonical, and diagonalizing the mass matrix, the result is the coupling in \cref{L} \cite{Hook:2010tw}. Here we used that $\eps \ll1$ and have dropped the $\O(\eps^2)$ terms, and we have neglected the coupling to the $Z$-boson, valid if the dark photon mass is small compared to the electroweak scale.

Finally, the last term is the WZW action, which is present if the 5th homotopy group of the coset space $\pi_5(G/H)$ is non-trivial; this is the case for $N_f \geq 3$ mass degenerate flavors.

Expanding the action in pion fields, the Lagrangian is the sum of the Lagrangian for chiral perturbation theory ($\chi$PT), the dark photon Lagrangian, and the terms from the WZW action: $\L = L_{\chi {\rm PT}} + \L_{V}+\L_{\rm WZW}$. The relevant dark pion and and dark photon interactions are:
\begin{align}
\L_{\chi {\rm PT}}  &= \Tr (\partial \vecS \pi)^2 -m_\pi^2 \Tr  (\vecS  \pi^2)
+\frac{1}{3f_\pi^2}\Tr \( (2\vecS \pi \partial \vecS \pi)
(\vecS \pi \partial \vecS \pi) -2(\vecS \pi \vecS \pi)( \partial \vecS
\pi\partial\vecS \pi)+m_\pi^2 \vecS \pi^4\) \nn \\
& + 2ig_dV^\mu     \Tr \(( \partial_\mu \vecS \pi) [Q,\vecS
     \pi]\), \nn \\
 \L_{V}  &= -\frac14 V_{\mu\nu}^2 - \frac12 m_V V_\mu^2 - \eps V_\mu \sum_f q_f \bar f \gamma^\mu f \nn \\
 \L_{\rm WZW}&=\frac{2N_c}{15 \pi^2 f_\pi^5} \eps^{\mu\nu\rho
  \sigma} \Tr\[ \vecS \pi \partial_\mu \vecS\pi \partial_\nu \vecS\pi \partial_\rho
  \vecS\pi \partial_\sigma \vecS \pi\]
                 -i\frac{N_c g_d}{3\pi^2 f_\pi^3}\eps^{\mu\nu\rho\sigma} V_\mu \Tr \( Q \partial_\mu \vecS \pi \partial_\rho \vecS \pi \partial_\sigma \vecS \pi\),
                 \nn \\ & +\frac{N_c g_d^2}{4\pi^2 f_\pi} \eps^{\mu\nu\rho\sigma} 
      (\partial_\mu V_\nu) V_\rho 
\Tr \(
    Q^2  \partial_\sigma \vecS \pi \).
 \label{L_pi}
\end{align}
The first two terms in the chiral lagrangian are the kinetic and mass term for the pion fields, with mass $m_\pi^2 = 2 \zeta f_\pi m_q$. 
Chiral perturbation theory is perturbative for
\be
\xi \equiv \frac{m_\pi}{f_\pi} \lesssim 4\pi.
\label{perturbative}
\ee
The 3rd term gives the 4pnt pion interactions, and the last term  the pion-dark photon coupling ($V 2\pi$ interaction). The first two terms in the dark photon Lagrangian are  the kinetic and mass term for the dark photon field, and the last term the coupling of the dark photon to the SM fermions from kinetic mixing. Finally, the WZW lagrangian contains the 5pnt pion interaction, and additional dark photon-pion couplings with an odd number of pions ($V 3\pi$ and $2 V \pi$ interactions). We have only included the most relevant, lowest dimensional operators.

Higher dimensional operators scale with the inverse cutoff scale $1/f_\pi$, which is suppressed compared to the pion mass scale for small $\xi$. For odd $N_f$ these terms can make the neutral pion unstable \cite{Berlin:2018tvf}, which can be avoided by introducing a small mass splitting between the quarks \cite{Katz:2020ywn}. We will instead assume that for odd $N_f$ the Wilson coefficients of the higher order contributions are sufficiently small for the dark pion to be stable on cosmological timescales, and present results for the minimal setup with $N_f=3$ as well as for $N_f=4$.

\section{Self-interactions}
\label{sec:self-int}

The dark pion can scatter via a 4-point contact interaction appearing in the $\chi$PT Lagrangian, and via the exchange of a dark photon.  The cross sections for these contributions are calculated in the non-relativistic limit in \cref{A:self_interactions}. The photon exchange contribution is subdominant, unless enhanced by an $s$-channel resonance which can appear for fine-tuned dark photon masses \cref{mV}. The cross section is can then be approximated by a sum of the velocity independent contact interaction and velocity dependent resonance contribution $\sigma_{\rm SI} \approx  \sigma_{\rm SI} ^{4{\rm pnt}}+\sigma_{\rm SI} ^{\rm res}$.

For SIMP dark matter the self-interactions are naturally large, and the Bullet cluster observations \cite{Markevitch:2003at,Clowe:2006eq,Randall:2008ppe} put a strong constraint on the cross section.  In the resonant self-interacting dark matter (RSIDM) scenario \cite{Chu:2018fzy,Tsai:2020vpi},  with a judicial choice of parameters, the self-interactions can affect structure formation on small scales in a velocity dependent way, and thus may address the putative problems of collisionless dark matter \cite{Spergel:1999mh,deBlok:2009sp,Boylan-Kolchin:2011qkt,Weinberg:2013aya}.

\subsection{Bullet cluster bound}

The $s$-wave part of the dark pion self-interaction cross section is \cref{sigma_4pnt}
\be
\frac{\sigma_{\rm SI}^{4{\rm pnt}} }{m_\pi}= \frac{3\kappa_{\rm SI} \xi^4}{64\pi m_\pi^3}
= 
 2.2 \times 10^5 \,\frac{{\rm cm}^2}{{\rm g}}\(\frac{{\rm MeV}}{m_\pi}\)^3  \frac{ 3\xi^4 \kappa_{\rm SI} }{64 \pi } \leq a_{\rm int} \frac{{\rm cm}^2}{{\rm g}},
 \label{a_int}
\ee
with
$\kappa_{\rm SI} = ( N_f^4-\frac23
  N_f^2+2)/(N_f^2 (N_f^2-1)) = 1 +\O(N_f^{-2})      $
and ${\rm MeV}^{-3} \approx2.2 \times 10^5 \, {\rm cm}^2/{\rm g}$. The cross section is not very sensitive to the number of quark flavors $N_f$. 
The Bullet cluster observation puts an absolute upper bound on the self-interaction cross section, given by  $a_{\rm int} \approx 1$. In the RSIDM scenario, where the resonant interactions from dark photon exchange become important at small scales, the data is fit  by a smaller $s$-wave contribution and $a_{\rm int} \approx0.11$ (and the r.h.s. of \cref{a_int} becomes an equality)  \cite{Chu:2018fzy}. The bound on the cross section can be 
translated in a condition on the dark pion mass
\begin{align}
m_\pi&\geq 14.9 \, {\rm MeV} \(\frac{\xi^{4}\kappa_{\rm SI}}{a_{\rm int}}\)^{1/3}.
                        \label{mass_bound}
\end{align}
%

\subsection{Resonant self-interactions}

The velocity dependent contribution to the self-interactions from nearly on-shell dark photon exchange can be written in Breit-Wigner form \cref{self_res} 
\begin{align}
  \sigma^{\rm res}_{\rm SI} &=
           \frac{4\pi S}{m_\pi E(v)}
           \frac{ \Gamma_d(v)^2/4}{  (E(v)-E(v_R))^2 + {\Gamma(v)^2}/{4}},
\end{align}
with $S =3S_f/N_\pi^2$ the ratio of multiplicities of the resonance dark photon (3 polarizations) and DM particles ($N_\pi =N_f^2-1$) times a symmetry factor $S_f=2$ that takes into account that there are two identical particles in the final state of both the self-interaction process and in dark photon decay. $E(v) = m_\pi v^2 /4$ is the kinetic energy in terms of the relative velocity $v$, and $E(v_R) = m_\pi \delta m$ the resonant kinetic energy.  Further, $\Gamma_d(v)$ is the running decay rate of the dark photon into dark sector pions, and $\Gamma=\Gamma_d +\Gamma_v$ the total running decay rate, which includes the decay into the visible SM sector fermions and pions via kinetic mixing. The velocity dependence of the decay widths can be parameterized as $\Gamma_i = m_V \gamma_i v^{n_i}$ with $i=d,v$; explicitly (see \cref{A:decay})
\begin{align}
  \Gamma_d(v) = m_{V} \( \frac{C_4 \alpha_d}{24 S_f}\) v^3 , \qquad
  \Gamma_v(v) = m_{V} \( \frac{\alpha\eps^2}{3\pi S_f}\) v^0,
  \label{rates}
\end{align}
with $\alpha_d = g_d^2/(4\pi)$ and $\alpha =e^2/(4\pi)$ the dark sector and SM fine-structure constants. 

The resonance is peaked for $v =v_R$, with $v_R 
= 3.6 \times 10^{-4} c$ from small scale structure data \cite{Chu:2018fzy}.  This determines $\delta m$
\be
\delta m= \(\frac{v_R}{2}\)^2 = 3.2 \times 10^{-8}.
\label{dm_data}
\ee
The height of the peak is fit by $m_\pi = 4000\,{\rm MeV} S^{1/3} (B_d \gamma_d)^{1/3}$, with $B_d =\Gamma_d/\Gamma$ the branching ratio for decay into the dark sector. This determines the dark photon gauge coupling:
\be
\alpha_d = 1.3 \times 10^{-4} \(\frac{m_\pi}{100\,{\rm MeV}}\)^3 \frac{N_\pi^2 }{B_dC_4}
\label{alpha_bound}
\ee
The resonant enhancement of the cross section is large, and small $\alpha_d$ is required to avoid too large self-interactions. 

For RSIDM both the mass and the dark gauge coupling are given in terms of $\xi$, which together with the kinetic mixing parameter $\eps$ are the only free parameters left. The combined constraints \cref{mass_bound} with $a_{\rm int}=0.11$ and \cref{dm_data,alpha_bound} give
\begin{align}
  m_\pi= 31.0 \, {\rm MeV} \(\xi^{4}\kappa_{\rm SI}\)^{1/3},\quad
\alpha_d = 3.7 \times 10^{-6} \xi^4 \frac{N_\pi^2 \kappa_{\rm SI}}{C_4 B_d},\quad
\delta m= 3.2 \times 10^{-8}.
\label{resonant}
\end{align}
%

\section{Relic density }
\label{sec:freeze-out}

The dark pions can freeze out via 3 → 2 number changing SIMP interactions and via annihi- lation into SM fermions. In the SIMP scenario thermal equilibrium with the SM sector should be maintained through freeze-out, to avoid entropy production and heating up of the dark sector. We will assume this is the case in this section, and return to the question whether the dark photon can be responsible for this in the next section.

The SIMP interactions get a contribution form the 5-point coupling in the WZW Lagrangian, and from diagrams with dark photon exchange allowed by the $(3\pi)V$ coupling in the WZW Lagrangian; with our choice of dark charges \cref{Q} the $\pi(2V)$ coupling vanishes.

The thermally averaged cross section is calculated in \cref{A:SIMP}, and is given in \cref{alpha_eff,SIMP_photon}.
Introducing the `time' variable
\be
x = \frac{m_\pi}{T},
\label{x}
\ee
it can be written in the form
\begin{align}
  \langle \sigma v^2 \rangle_{3 \to 2}\approx \langle \sigma v^2 \rangle_{3 \to 2}^{5{\rm pnt}}+ \langle \sigma v^2 \rangle_{3 \to 2}^{\rm res}
  =\frac{\alpha_{3 \to 2}}{x^2 m_\pi^5}(1+ \alpha_{\rm res}x^{5/2}\e^{-\delta m\, x}).
  \label{sigma_32}
\end{align}
The first term comes from the 5pnt pointlike interaction. The 2nd term is from dark photon exchange, which is dominated by an $s$-channel resonance if the dark photon is nearly on shell $\delta m \ll 1$, and we have we used the narrow width approximation to evaluate the thermally averaged cross section. We have calculated the latter term only for $N_f =3$ and $N_f=4$ flavors.
 The effective couplings are 
\be
\alpha_{3 \to 2}=\frac{5\sqrt{5} }{1536 \pi^5}\frac{N_c^2 \kappa_{3 \to 2} \xi^{10}}{N_f}, \qquad
\alpha_{\rm res}|_{N_f=3,4} = \frac{512 \pi^{5/2}}{15C_4} \frac{\alpha_d}{\xi^4}
                  \label{alpha}
\ee
with $\kappa_{3 \to 2} =N_f^2 (N_f^2-4)/N_\pi^2 =  1 +\O(1/N_f)$. The resonant contribution dominates at freeze-out $x =x_f$ for $\alpha_d/\xi^4 \gtrsim 3.8 \times 10^{-6} x_{f 20}$ with $x_{f 20} = x_f/20$. For RSIDM \cref{resonant} this is the case for $\xi \lesssim {\cal O}(1)$. 

Dark pions can annihilate into SM electrons (and depending on their mass, into heavier charged SM particles) via the kinetic mixing portal. Annihilation is also dominated by the $s$-channel resonance and the thermally averaged cross section is \cref{sigma_ann_res}
\begin{align}
\langle \sigma v \rangle_{\rm ann}
    =\alpha_{\rm ann}\frac{x^{3/2}\e^{-\delta m \, x}}{m_\pi^2} ,\quad \alpha_{\rm ann} =\frac{32 \pi\sqrt{\pi} \eps^2 \alpha B_d}{N_\pi^2 }
  \label{alpha_ann}
\end{align}   
with as before $B_d = \Gamma_d/\Gamma$ the branching ratio for decay into dark sector states. We note that both the resonant part of the WZW interactions and the resonant annihilations are independent of the mass splitting $\delta m$, except for the exponential factor, which determines below which temperature $x \gtrsim 1/\delta m$ these interactions are  `turned off'.

The Boltzmann equation for the dark pions reads
\be
\dot n +3 H n = -\langle\sigma v\rangle_{\rm ann} (n^2 -n_{\rm eq}^2 ) - \langle\sigma v^2\rangle_{3\to2}(n^3 -n^2 n_{\rm eq} ),
\label{Boltz0}
\ee
which in terms of the number density fraction $Y=n/s$ becomes
\be
\frac{\dd Y}{\dd x} = - \frac{\lambda_{3\to 2}}{x^7}( 1+\alpha_{\rm res}x^{5/2}\e^{-\delta m\, x}) (Y^3-Y_{\rm eq} Y^2)-\frac{\lambda_{\rm ann}\e^{-\delta m\, x}}{\sqrt{x}} (Y^2-Y_{\rm eq}^2).
\label{Boltz1}
\ee
Here
\be
\lambda_{3\to2} = \frac{s(m_\pi)^2 \alpha_{3 \to 2}}{m_\pi^5 H(m_\pi)},\quad
\lambda_{\rm ann}=\frac{ s(m_\pi) \alpha_{\rm ann}}{m_\pi^2 H(m_\pi)}
\ee
with $ s(m_\pi) = (2\pi^2 g_{*s}m_\pi^3)/45$ and $ H(m_\pi) = (\pi \sqrt{g_*} m_\pi^2)/(3\sqrt{10}\mpl)$ the entropy density and Hubble constant at  $T =m_\pi$.

The relic dark matter density matches observations \cite{Planck:2018vyg} for
%

\be
\Omega_{\pi,0} = \frac{N_\pi m_\pi s_0 Y_\infty}{\rho_c}
\quad
\Rightarrow \quad m_\pi Y_\infty 
=0.4 \times 10^{-6} \,{\rm MeV}
\label{Omega}
\ee
with $Y_\infty= n/s$ the asymptotic  number density fraction, and $s_0$ the entropy density today.

\subsection{SIMP freeze-out}
\label{s:freeze-out}

Consider first the case that freeze out of the dark pion is determined by the WZW interactions, either by the 5pnt contact interaction or the dark photon mediated contribution.  This means annihilation are subdominant at freeze out $\langle\sigma v\rangle_{\rm ann} \lesssim \langle\sigma v^2\rangle_{3\to2} \, n_{\rm eq}$ at $x=x_f$. We will estimate the bound this condition gives on $\eps$ in the next section.  Note, however, that the annihilation cross section grows at late times, and even if negligible at freeze out, annihilation may still affect the final relic density significantly.

To describe freeze out we thus set the annihilation contribution to zero in the Boltzmann \cref{Boltz1}. At late times the equilibrium distributions can be dropped, which allows to solve for the asymptotic distribution
\be
\lim_{x \to \infty}\frac{\dd Y}{\dd x} =- \frac{\lambda_{3\to 2}}{x^{7}}(1+ \alpha_{\mathrm{res}}x^{5/2})Y^3 \quad \Rightarrow \quad Y_f \simeq \frac{\sqrt{ 3}x_f^3}{\sqrt{\lambda_{3\rightarrow2}}}\frac{1}{\sqrt{1+\frac{12}7 \alpha_{\mathrm{res}}x_f^{5/2}}}
\label{Yf_SIMP}
\ee
where we introduced the notation $Y_f$ for the asymptotic number density after freeze-out of the SIMP reactions.  The freeze-out temperature can be estimated from $n_\pi^2 \langle \sigma v^2 \rangle_{3\to 2} \simeq H$ which gives 
\begin{align}
  x^{3}\e^{2x} \big |_{x=x_f}\simeq
  \frac{N_\pi^2 \alpha_{3 \to 2}
  m_\pi }{(2\pi)^3 H(m_\pi) }(1+ \alpha_{\mathrm{res}}x^{5/2}) \equiv C_{3\to 2} (1+ \alpha_{\mathrm{res}}x^{5/2})
  \label{xf}
\end{align}
where we used that the non-relativistic number density is $n= N_\pi m^3_\pi(2\pi x)^{-3/2}\e^{-x}$. In the limit that the 5pnt interaction respectively the resonant contribution dominates the cross section we can estimate the freeze-out temperature using that $x^n e^{2x} =c$ gives $x \approx \ln \sqrt{c} -\frac{n}{2} \ln(\ln\sqrt{c})$.

The SIMP interactions are negligible after freeze out, but annihilations may still have an effect. To estimate this we solve the Boltzmann equation with the boundary condition $Y(x_f)=Y_f$:
\be
\frac{\dd Y}{\dd x} = -\frac{\lambda_{\rm ann}\e^{-\delta m\, x}}{\sqrt{x}} Y^2, \qquad \Rightarrow \qquad
Y_\infty
\approx Y_f  \frac{\sqrt{\delta m} }{\sqrt{\delta m} + \sqrt{\pi} Y_f \lambda_{\rm ann}}
\label{Y_infty}
\ee
where we assumed that $(\delta  m \,x_f) \ll 1$. The annihilation rate increases at late time,  and is most efficient just before the exponential cutoff at $x = 1/\delta m$ kicks in; this is how the $\delta m$-dependence appears in the estimate for the relic density.
It follows that annihilations are negligible if
\be
\frac{\sqrt{\pi} Y_f \lambda_{\rm ann}}{\sqrt{\delta m}} < 1  \qquad \Rightarrow \qquad
\eps <7.6 \times 10^{-9} N_\pi\sqrt{B_d \delta m}\(\frac{m_\pi}{{\rm MeV}} \).
\label{ann_important}
\ee
that is, only for very small kinetic mixing.

\subsection{Annihilation scenario}

In the opposite limit that $3\to2$ interactions are always subdominant, $\langle\sigma v\rangle_{\rm ann} \ge \langle\sigma v^2\rangle_{3\to2} \, n_{\rm eq}$ at freeze-out, the relic density is set by annihilation reactions only. We can still use \cref{Y_infty} for the relic density, but now $Y_f(x_f)$ is the number density as annihilations freeze out.
The freeze-out temperature can be estimated from $n_\pi \langle \sigma v\rangle_{\rm ann}
\simeq H$ which gives
\begin{align}
  x^{-2}\e^{x} \simeq
  \frac{N_\pi \alpha_{\rm
  ann} m_\pi }{(2\pi)^{3/2} H(m_\pi) } \equiv C_{\rm ann} \qquad \Rightarrow \qquad x_f \simeq
  \ln C_{\rm ann} +2 \ln (\ln C_{\rm ann}).
  \label{xf_ann}
\end{align}
We estimate the freeze out density
\be
Y_f \approx \frac{n}{s}\Big|_{x_f}= \frac{x_f^{7/2}}{\lambda_{\rm ann}}
\label{Yf_ann}
\ee
where we used \cref{xf_ann}. If $x_f$ in \cref{Yf_ann} is smaller than that for SIMP reactions \cref{xf}, it follows freeze-out is dominated by annihilations. An earlier freeze out means a larger density $Y_f$.
Hence we can write the asymptotic number density as 
\be
Y_\infty
\approx Y_f  \frac{\sqrt{\delta m} }{\sqrt{\delta m} + \sqrt{\pi} Y_f \lambda_{\rm ann}}, \qquad
Y_f ={\rm max}\(Y_f \big|_{\rm ann},\, Y_f \big|_{3\to2}\)
\label{Y_infty_all}
\ee
with the freeze out density for $3 \to2$ reactions and annihilation  given in \cref{xf,Yf_SIMP}, and \cref{xf_ann,Yf_ann} respectively.

\section{Kinetic mixing}
\label{sec:mixing}
Kinetic mixing between the dark and SM photons provides a portal between the dark and visible sector. In the SIMP scenario, in which the relic density is determined by the freeze-out of $3\to2$ dark pion number changing interactions, both sectors need to be in thermal equilibrium during freeze-out to avoid heating up the dark sector.  The SIMP scenario further requires that dark pion annihilation into SM particles is subdominant during freeze-out -- although, as we have seen in the previous subsection, annihilation still may affect the relic density at late times. In this subsection we determine the constraints on the mixing parameter $\eps$ that these two requirements give. The relevant cross sections are computed in \cref{A:dark_photon}. We also quickly review the relevant cosmological and collider bounds on the kinetic mixing parameter.

\subsection{Thermal equilibrium between the dark and visible sector}

The dark and visible sector can be kept in thermal equilibrium via pion scattering with SM electrons and positrons.
The non-relativistic cross section for this process is \cref{sigma_pi_SM}
\begin{align}
\sigma_{\rm scat} 
  = A _{\rm scat} \eps^2\frac{ p^2}{m_{\pi}^4},  \qquad A_{\rm scat} = \frac{2 \pi C_4 \alpha_D \alpha }{N_\pi}
\end{align}
This can be straightforwardly generalized to include muon and SM pion scattering as well,\footnote{Scattering off muons was included in the numerical results, but its effect for thermalisation was found to be negligible.} see below \cref{sigma_pi_SM}, in case freeze-out occurs at temperatures exceeding the muon and pion mass. Here $p \approx E_e$ is the incoming electron momentum, and $C_4 =4\, (8)$ for $N_f =3\,(4)$ the same color factor as appearing in the dark photon decay rate \cref{rates}.  The scattering rate can be estimated as $\Gamma_{\rm scat} \approx\langle n_e E_e^2 \rangle {(\sigma v)_{\rm scat} }/{E_e^2}$ \cite{Hochberg:2015vrg}, and demanding that it exceeds the Hubble rate $\Gamma_{\rm scat} > H$ at the time of freeze out gives the bound
\begin{align}
\eps > \sqrt{\frac{H(T_f) m_\pi^4}{\langle n_e E_e^2 \rangle  A _{\rm scat} }} 
  &    = 4.6 \times 10^{-8} x_{f20} ^{3/2}\sqrt{\frac{m_\pi}{100\,{\rm MeV}}\frac{4}{g_e} \frac{N_\pi}{C_4\alpha_d}} ,
    \label{eq_kin}
\end{align}
where we used $\langle n_e E_e^2 \rangle |_{x=x_f}=\frac{g_e 45 \zeta(5) m_\pi^5}{4\pi^2 x_f^5}$, $H(T) =H(m_\pi)/x_f^2$, and as before $x_{f20} =x_f/20$. Further, $g_e=4$ are the degrees of freedom of the electron/positron pair. 
To get the numerical value we used $\alpha =1/137$ and $g_s =17.25$. For RSIDM  the bounds \cref{resonant} eliminate the dark pion mass and gauge coupling dependence. 

If the dark photon is to maintain thermal equilibrium with the SM, annihilations cannot be neglected for the relic density calculation if (comparing eq.~\ref{ann_important} and eq.~\ref{eq_kin})
\be
\label{eps_RSIDM}
m_\pi \gtrsim\frac{ 3.7 \times 10^{-2}\ \mathrm{MeV}}{\alpha_d \delta m}\ \frac{10}{C_4 N_\pi } x_{f20}^3 \overset{(\ref{resonant})}{=} 3600\ \mathrm{MeV}\ \Big( \frac{8 x_{f20}}{N_\pi} \Big)^{3/4}.
\ee
The last equality applies to RSIDM, for which annihilations thus always play a role, depleting the dark matter abundance at late times.


\subsection{Annihilations subdominant during freeze-out}

Annihilations are subdominant at freeze-out if  $\langle\sigma v\rangle_{\rm ann}  < \langle\sigma v^2\rangle_{3\to2} \, n_{\pi}$ at $x=x_f$ \cref{xf}, which translates to $ \sqrt{\alpha_{3 \to 2}(1+\alpha_{\rm res}x_f^{5/2})H(m_\pi)/m_\pi} > \alpha_{\rm ann} x^{7/2}_f$. This gives an upper bound on the kinetic mixing parameter
\begin{align}
  \eps& < 4 \times 10^{-8}  \( \frac{m_\pi}{100\,{\rm MeV}}\)^{1/4}\frac{\xi^{5/2}}{x_{f20}^{7/4} } \frac{N_\pi \sqrt{N_c}}{N_f^{1/4} } \Big(1+3\times 10^5 \frac{\alpha_d x_{f20}^{5/2}}{\xi^4}\Big)^{1/4},
        \label{eps_subdom}
\end{align}
where we set $\kappa_{3 \to 2}, B_d$ to unity.

%

\subsection{CMB bound and other bounds on kinetic mixing}

BBN and CMB observations place bounds on the energy injected in the photon fluid from late time dark pion annihilations to SM final states. These constraints can be stringent as the resonant contribution to the thermally averaged cross section grows as $\langle \sigma v \rangle \propto x^{3/2} e^{-\delta m x}$. At late times $x > \delta m^{-1}$, when the (average) dark matter velocity falls below the resonance velocity, resonant annihilations into SM final states are exponentially suppressed. Due to the momentum dependent coupling of the dark pions to the dark photon the remaining non-resonant contribution is $p$-wave suppressed, and evades all CMB bounds. For self-interacting resonant DM the mass splitting $\delta m \sim 3\times 10^{-8}$ is small \cref{dm_data}, and the exponential suppression of the cross section only kicks in shortly before or after the CMB is formed, depending on the mass of the dark pions. The thermally averaged cross section thus peaks at this time. CMB observations are more stringent for $s$-wave annihilations (as opposed to more stringent BBN bounds for $p$-wave annihilations) \cite{Depta:2019lbe}. Since our thermally averaged cross sections scales inversely proportional to the velocity, CMB bounds are stronger than bounds from BBN, and we thus only consider the former.

The CMB bound is given by $p_{\rm ann} < 3.3 \times 10^{-31}\ \mathrm{cm^3 s^{-1} MeV^{-1}}$, where $p_{\rm ann} \equiv f(z) \frac{\langle \sigma v\rangle}{m_\pi}$ and $f(z) = 0.01-1$ is a function that quantifies the efficiency of energy injection in the CMB \cite{Planck:2018vyg}. Recasting this to a bound on the mixing parameter $\epsilon$, the constraint is given by (setting $f(z)=1$)
\be
  \epsilon \lesssim 9.4 \times 10^{-14} \Big( \frac{m_\pi}{10\ \mathrm{MeV}}\Big)^{3/4} \exp{\Big(1.1 \frac{m_\pi}{10\ \mathrm{MeV}}\Big)},
  \label{CMB}
  \ee
which vanishes above dark pion masses of 150 MeV. This bound was derived for s-wave scattering. In our model, the annihilation cross section increases for later times until the exponential cut-off kicks in, so the bound is expected to be stronger at late times. At the same time, the energy injection into the CMB is maximized at $z\sim 600$, where $f(z)\approx 1$ \cite{Slatyer:2015jla,Slatyer:2015kla}.  
Although applying the CMB bound naively for our case at the different choices of $z$ affects the exact constraint on the mixing parameter somewhat, this has no consequences on parameter space of the SIMP scenarios discussed in the next section, as this is dominated by the constraints from thermalisation and beam dump experiments. For simplicity, then, we imposed the CMB bound at $z\sim 600$, and that is what is shown in our figures in \cref{sec:scenarios}.
  
For larger $\delta m$ only the BBN bound applies. 
Following \cite{Depta:2019lbe}, energy injection during BBN is most efficient in the range $1/T \sim 10^2-10^3\ \mathrm{MeV}^{-1}$. Requiring the annihilations to be suppressed at this time ($\delta m x \gtrsim 1$), the BBN bound can evaded for mass splittings $\delta m  \gtrsim 10^{-3}$. 

Finally, there are bounds from dark photon searches at beam dump or fixed target experiments at electron or proton colliders. In these experiments large number of dark photons can be produced from Bremsstrahlung or secondary meson decays. The experiments typically search for highly displaced vertices in the detector \cite{Feng:2022inv,Fabbrichesi:2020wbt}. For our region of interest, $10^{-4}\lesssim \epsilon \lesssim 10^{-8}$ and $10\ \mathrm{MeV} \lesssim m_{\gamma'} \lesssim 1\ \mathrm{GeV}$ we consider bounds from NuCal\cite{Blumlein:2011mv,Blumlein:2013cua}, CHARM\cite{CHARM:1985nku}, and E137 \cite{Bjorken:1988as}. Given the $p$-wave nature of our scattering cross section, scattering at late times is heavily suppressed. Bounds on millicharged particles from direct detection experiments like XENON are therefore too weak to constrain the kinetic mixing parameters and are therefore not considered.

\section{SIMP scenarios}
\label{sec:scenarios}

In the `standard' SIMP scenario the dark pion relic density results from the freeze out of the 5pnt $3\to 2$ WZW processes.  We reproduce this set-up by turning off the dark photon interactions $\alpha_d =0$.  The Bullet cluster observation puts an upper bound on the pion self-interactions, and consequently on the pion mass \cref{mass_bound}.  The correct relic density is obtained for $\xi\sim 4\pi$ for $N_f=N_c=3$,  uncomfortably close to the perturbativity bound $\xi \lesssim 4\pi$ \cite{Hochberg:2014kqa}. Increasing the number of colors improves the situation slightly, but a large number of colors is needed to be  within the perturbative regime.

\begin{figure}
    \centering
    \includegraphics[width=0.75\linewidth]{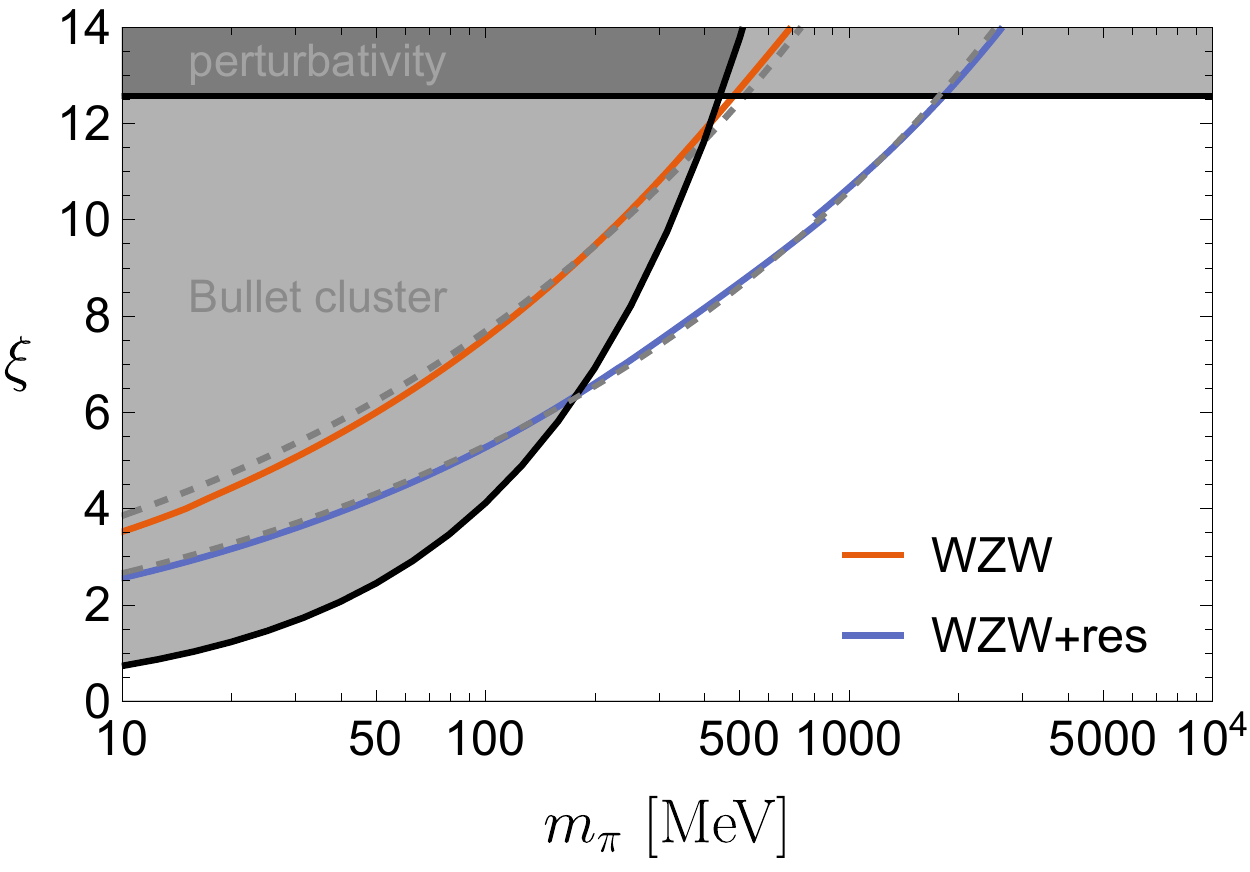}
    \caption{Relic density constraint on $\xi$ for different values of the dark pion mass for the WZW-term only (red), and including the resonance via the dark photon (blue). The dashed lines correspond to the estimate of \cref{Yf_SIMP} (setting $\alpha_d =0$ for the WZW estimate). The gray shaded areas are excluded by $\xi > 4\pi$, where the $\chi$PT description breaks down, and the Bullet cluster bound on the DM self-interaction $\sigma/m \leq 1\ \rm{cm^2/g}$. All lines are for $N_f=N_c=3$.}
    \label{fig:relicDensity}
\end{figure}

The problem with only the $3\to2$ interactions, and no resonance enhancement, is that dark matter is over produced.  The dark pions contribute a fraction to the DM density \cref{Omega}
\be
R \equiv \frac{\Omega_\pi}{\Omega_{\rm DM}}
\approx  \frac{9 \times 10^2 x_{f20}^3}{\xi^3N_c} \sqrt{\frac{N_f}{ a_{\rm int}}}.
\label{fractionR}
\ee
The freeze-out temperature only depends logarithmicly on the model parameters, and ranges from $x_{f20} = 0.68-1.1$ for $\xi=1-10$. The relic density is reproduced for $R=1$, which requires large $\xi \gtrsim 4\pi$. 

This is illustrated in \cref{fig:relicDensity}, which shows the numerical solution to the Boltzmann equation without (red) and with (blue) adding the $3\rightarrow2$ resonance, where $\alpha_d$ is chosen from \cref{resonant}. The dashed lines correspond to the estimate of \cref{Yf_SIMP} with $\alpha_{\rm res}=0$ ($\alpha_{\rm res}\neq0$) for the red (blue) curve. The shaded areas are excluded by the perturbativity cutoff on $\xi$ and the Bullet cluster bound on the self-interaction.

Without the dark photon resonance, the required value of $\xi$ is at or above the perturbativity cutoff. Including the resonance, but not considering annihilations, the situation improves significantly as the increased $3\to2$ interactions reduce the dark matter density. The observed relic density is obtained for a larger dark pion mass for a fixed $\xi$. For such a large pion mass, however, the kinetic mixing parameter is too small to maintain kinetic equilibrium with the SM \cref{eps_RSIDM} and either an additional portal interaction is required or annihilations should be considered. 

In the following subsections we discuss two SIMP scenarios that give the correct relic density, satisfy self-interaction constraints, and the kinetic mixing portal interaction maintains thermal equilibrium with the SM during freeze-out. 
First, we focus on the possibility that the dark pions are RSIDM. The $3\to2$ freeze-out interactions can be resonantly enhanced for large enough $\alpha_d/\xi^4 \gtrsim 10^{-6}$. Given the small mass difference $\delta m$ in \cref{dm_data},  annihilations become important at late times, and  reduce the relic density further. 
Second, we consider the more classical SIMP scenario, and drop the requirement that the self-interactions can affect small scale structure formation. The self-interactions should still satisfy the upper bound from the bullet cluster.
Both $3\to 2$ interactions and annihilations may be resonantly enhanced, by tuning the dark photon mass, but now have more freedom in the resonant condition $\delta m$ and the dark gauge coupling $\alpha_d$ to satisfy all constraints. 

\subsection{Self-interacting resonant SIMP DM}

Consider first the RSIDM scenario that the relic density is produced via the SIMP mechanism, i.e. via the freeze out of $3 \to 2$ reactions, and that resonant DM self-interactions can address the small scale structure problems.  The requirements on the self-interactions \cref{resonant} fixes the parameters $m_\pi, \alpha_d,\delta m$ in terms of $\xi$, which in turn is determined from the correct relic density \cref{xf,Yf_SIMP,Omega}.

For large enough kinetic mixing the relic density will be reduced by (late-time) annihilations, which reduces the required $\xi$ value. Assuming annihilations are subdominant during freeze-out and $B_d \approx 1$, then $Y_\infty$ is given by \cref{Y_infty} and $R$ now becomes
\be
R = \frac{2.7 \times 10^3 x_{f20}^3}{\xi^3} \frac{1}{R_{\rm res}+ 3.0 \times 10^{18} x_{f20}^3 \eps^2/ \xi^{17/3}}, 
\ee
where $R_{\rm res}\equiv \sqrt{1+30x_{f20}^{5/2}}$ accounts for the resonant enhancement of the WZW interactions, and the $\eps$-dependent term for late-time annihilations. Annihilations significantly reduce $R$ for $\eps > 7.6 \times 10^{-12} \xi^{17/6} x_{f20}^{-3/2} $ for sufficiently small $\xi \lesssim \O(1)$, but rapidly shut off for larger $\xi$. Solving for the observed relic density $R=1$ gives
\be
\label{eq:epsAnn}
\eps = \sqrt{9.3 \times 10^{-16} x_{f20}^3 \xi^{8/3} -  3.4 \times 10^{-19} R_{\rm res}\xi^{17/3}}\, x_{f20}^{-3/2}.
\ee

\begin{figure}
    \centering
    \includegraphics[width=0.75\linewidth]{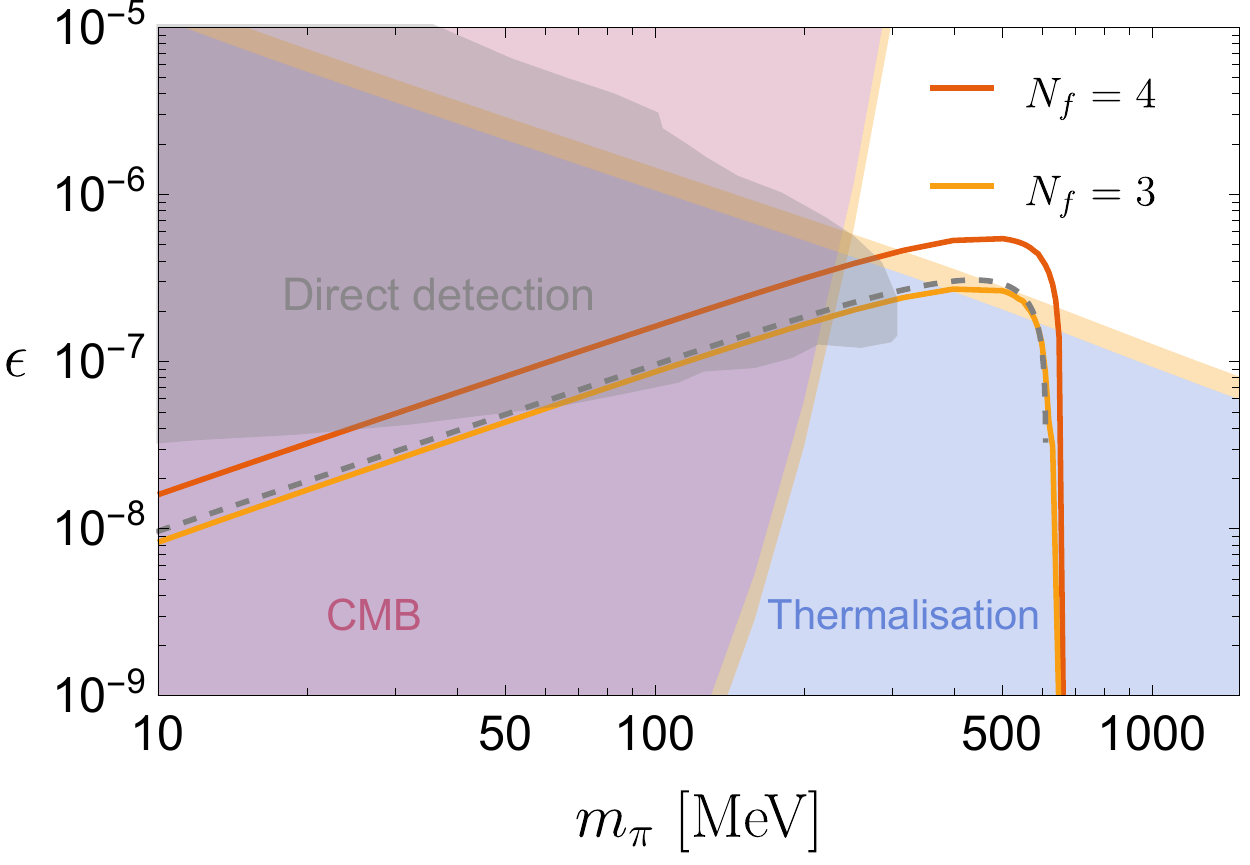}
    \caption{Constraints on the kinetic mixing parameter and dark pion mass in the  resonant SIDM scenario. The coloured (numerically) and dashed (estimate~\cref{eq:epsAnn}) lines show the values of $\epsilon$ for which the correct DM relic density is produced, and for which the dark pions have the required self-interaction. The blue shaded area excludes those values of $\epsilon$ for which the dark photon cannot maintain kinetic equilibrium with the SM. The purple shaded area depicts the CMB constraint on DM annihilations. The grey shaded area is excluded from beam dump searches from E137, nuCAL and CHARM. The exclusion bounds are slightly stronger for $N_f=3$ (indicated by the orange shaded areas) as the dark charge is smaller for $N_f=3$ (\cref{resonant}).}
    \label{fig:epsFull}
  \end{figure}

\Cref{fig:epsFull} shows the value of $\eps$ as a function of the dark pion mass $m_\pi$ for which $R=1$ from numerically solving the Boltzmann equations for $N_f=3$ (orange) and $N_f=4$ (red), as well as the analytical estimate \cref{eq:epsAnn} above (dashed curve). In addition, the bounds from the CMB, colliders, and from the requirement of thermal equilibrium between the  dark and visible sector during SIMP freeze out \cref{eq_kin} are shown.

The observed relic density can be obtained with perturbative couplings, e.g. $\xi=1$ for kinetic mixing $\eps = 3 \times 10^{-8}$. Note, however, that there is a maximum value
\be
\eps \le 2.9 \times 10^{-7}
\label{upper_eps}
\ee
to get the observed relic density, at a dark pion mass $m_\pi \sim 600-650$ MeV. For larger dark pion masses, the $3\rightarrow2$ interactions alone underproduce DM given the imposed relations on the self-interaction \cref{resonant}, so additional annihilations are not of any help; this explains the sharp cutoff of the red curve at large pion masses.  For such small kinetic mixing annihilations are subdominant during SIMP freeze-out \cref{eps_subdom}. For all dark pion masses of interest the dark photon decays predominantly into dark sector pions $B_d \approx 1$ \cref{eq:Bd}, validating our assumptions for the freeze out calculation.

The thermalisation requirement rules out most of the parameter space. Because the annihilations are highly efficient, and only a small portion of the DM should be depleted, the annihilation cross section should be suppressed by small values of $\eps$.  For $N_f=4$ and pion masses larger than $m_\pi \gtrsim 200\,$MeV the kinetic mixing is large enough to maintain kinetic equilibrium.
For $N_f=3$, on the other hand,  the constraints are slightly stronger as the dark charge is smaller \cref{resonant} -- indicated by the orange shaded areas in \Cref{fig:epsFull} -- and  the heat transfer to the SM is barely fast enough to maintain thermal equilibrium.

All constraints can be satisfied and the RSIDM scenario can be viable for $N_f=4$ and dark pion masses in the range $m_\pi\simeq  250-600\,$MeV, while for $N_f=3$ this is only marginally possible around $m_\pi \sim 500\,$MeV.  In our analysis we have analytically estimated the bound  on the required kinetic mixing with the SM. To make more precise statements requires inclusion of the bath effects, which also allows for (partial) heating of the dark bath, in our numerical calculations.
This is left for future research.




\subsection{SIMP DM}

We now consider the SIMP scenario, in which the relic density is determined by $3\rightarrow2$ interactions and possibly additional annihilations, but self-interactions are too weak to affect small scale structure formation.  As we have seen \cref{fractionR}, with just the 5pnt WZW interactions and given the Bullet cluster bound, too much DM is produced for perturbative couplings $\xi$. The DM density can be reduced by a resonant enhancement of the WZW interactions and by annihilations. No longer constrained by the small scale structure data, the value of $\delta m$ can now be larger. This immediately avoids CMB and BBN constraints as the thermally averaged annihilation cross section $\propto \e^{-\delta m\,x}$ \cref{alpha_ann} is exponentially suppressed in these eras. Moreover, for larger $\delta m$ dark photons will predominantly decay to dark pions, thus evading dark photon searches at beam dump experiments. For concreteness we will fix the mass splitting to $\delta m=10^{-3}$ throughout this subsection. This choice avoids the cosmological constraints, while it can still give rise to interesting phenomenology at late times.

In the parameter space region where the dark photon maintains thermal equilibrium with the SM sector during freeze-out of the WZW-interactions, the kinetic mixing parameter is always large enough that late time -- after freeze-out -- annihilations cannot be neglected. For general $m_\pi$, $\xi$, $\delta m$ and $\alpha_d$ the required mixing parameter that reproduces the correct relic density is
\be
\label{epsilon_full}
\eps = 1.5\times10^{-12} N_\pi \sqrt{\frac{\sqrt{\delta m}}{Y_f}\(\frac{m_\pi}{\mathrm{MeV}}\) \(3.8\times10^7 \(\frac{m_\pi}{\mathrm{MeV}}\)  Y_f -15\)},
\ee
with $Y_f$ from \cref{Yf_SIMP}. Parameter space opens up significantly compared to the RSIDM scenario, as all other parameters are free except for the Bullet cluster bound on the dark matter self-interactions.

\begin{figure}
\centering
\begin{subfigure}{0.48\textwidth}
    \includegraphics[width=\textwidth]{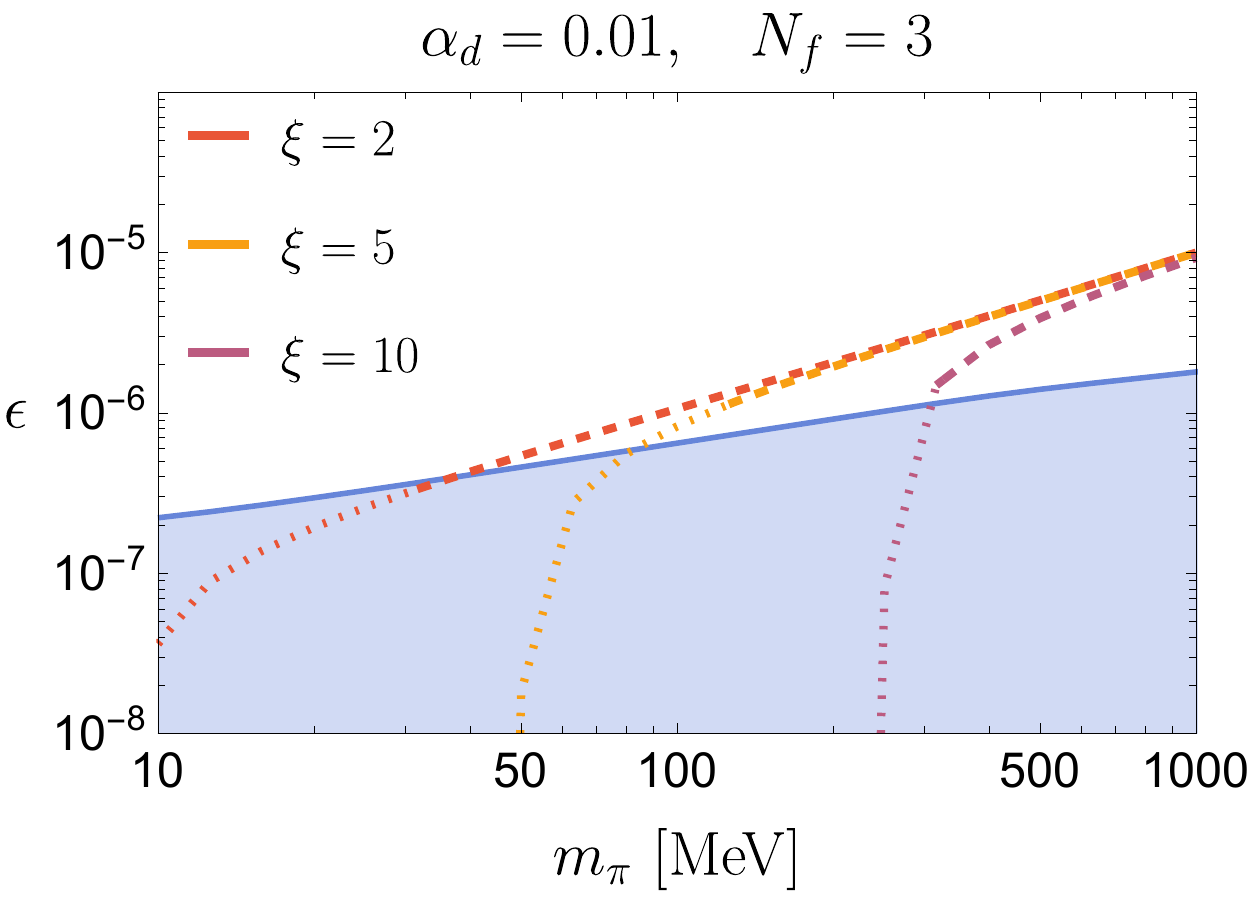}
\end{subfigure} 
\hfill
\begin{subfigure}{0.48\textwidth}
    \includegraphics[width=\textwidth]{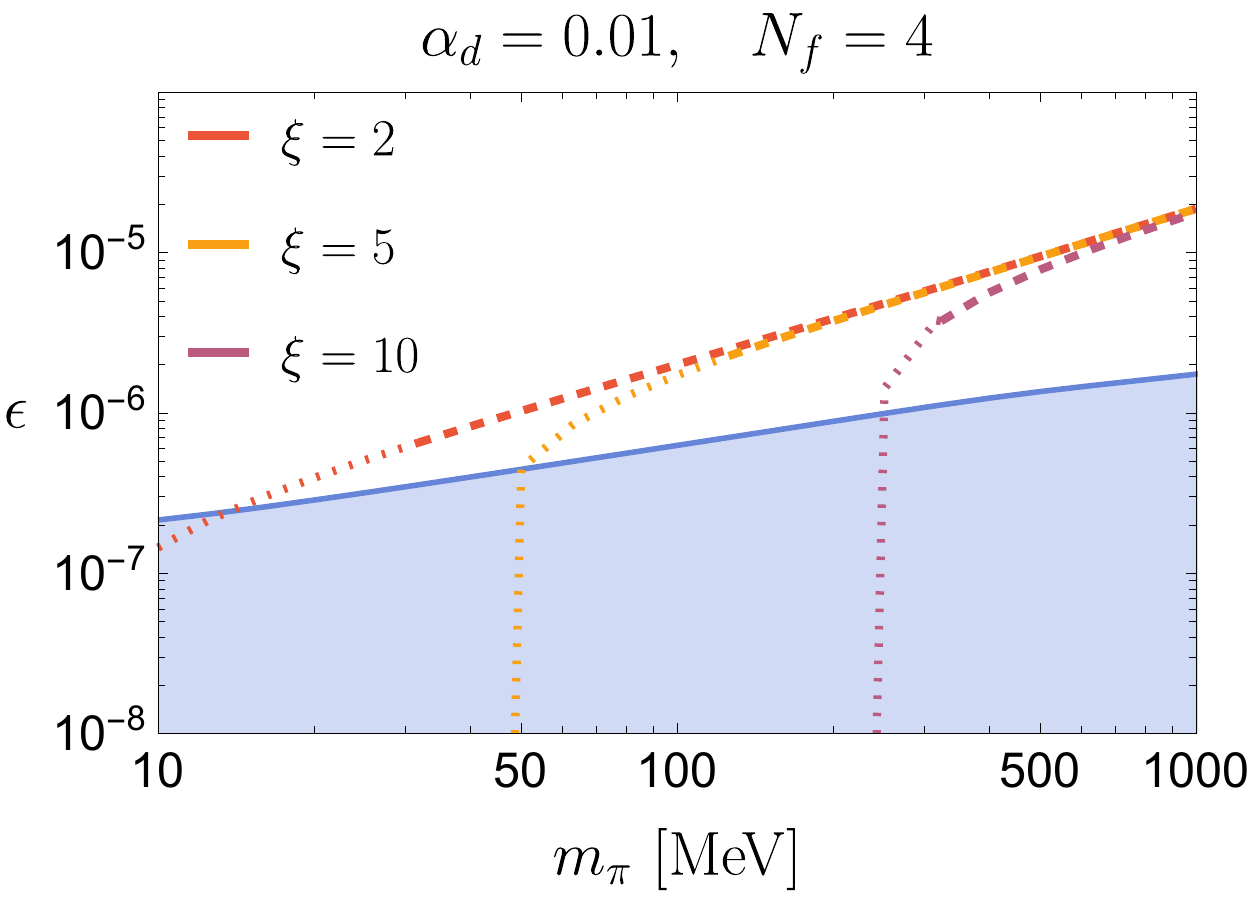}
\end{subfigure}
\hfill
\begin{subfigure}{0.48\textwidth}
    \includegraphics[width=\textwidth]{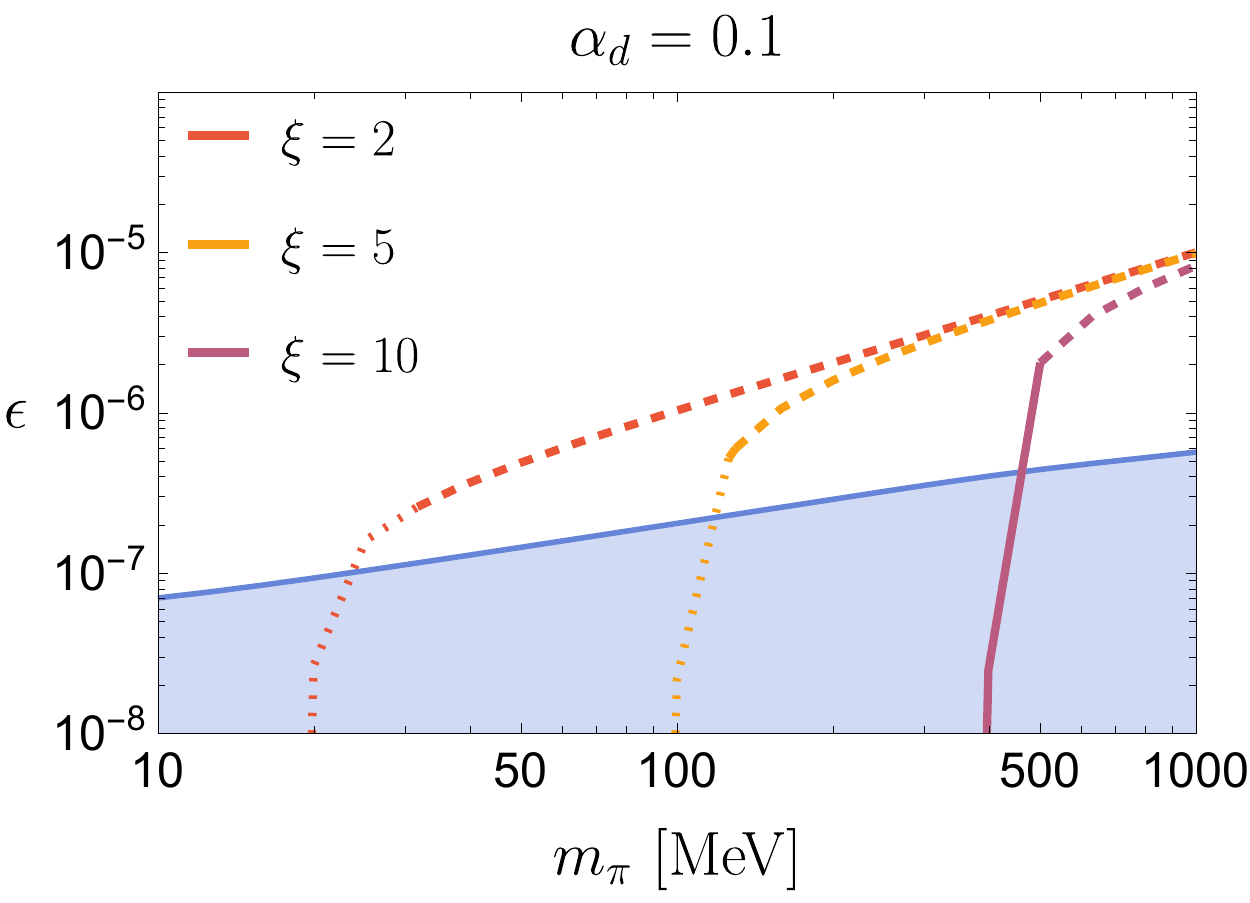}
\end{subfigure}
\hfill
\begin{subfigure}{0.48\textwidth}
    \includegraphics[width=\textwidth]{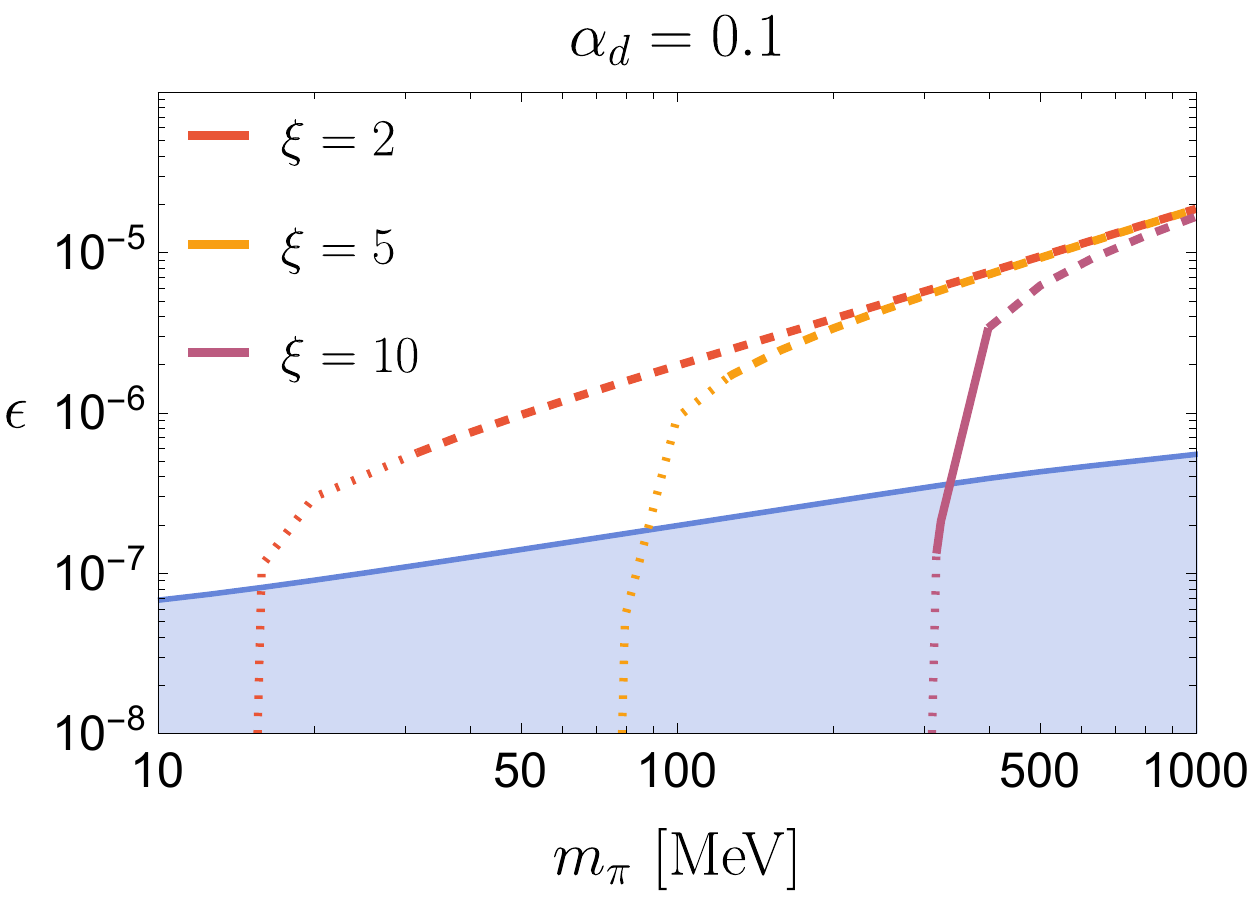}
\end{subfigure}
\hfill
\begin{subfigure}{0.48\textwidth}
    \includegraphics[width=\textwidth]{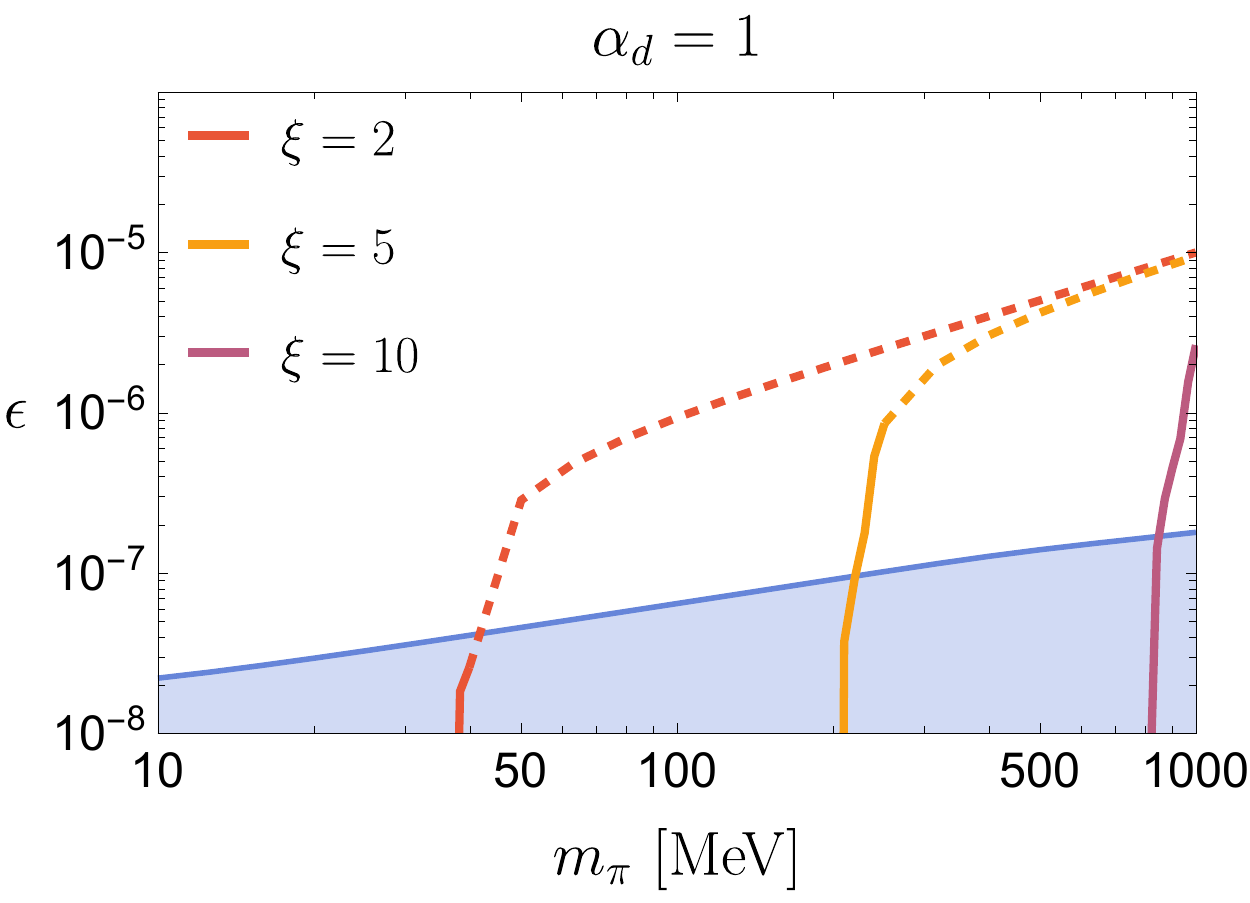}
\end{subfigure}
\hfill
\begin{subfigure}{0.48\textwidth}
    \includegraphics[width=\textwidth]{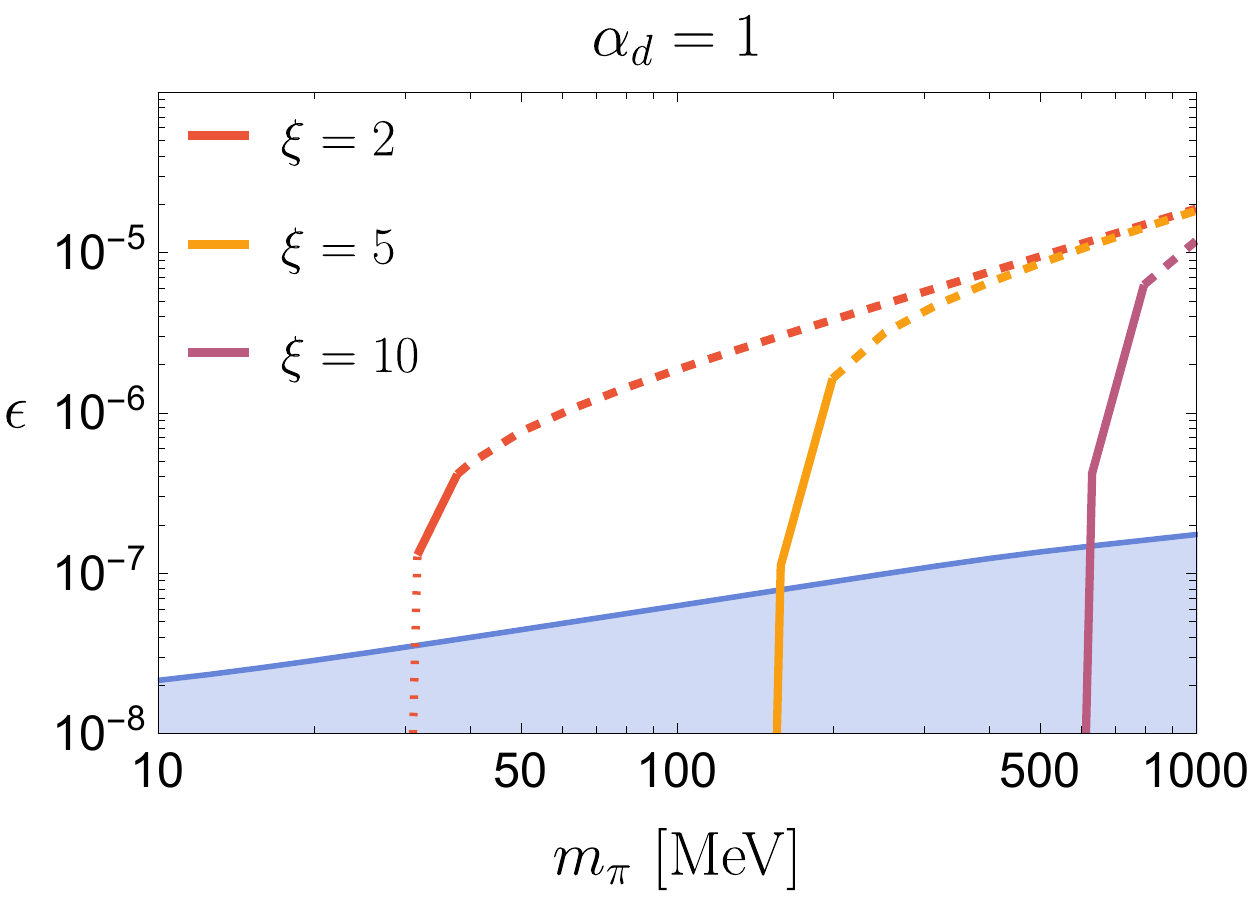}
\end{subfigure}
\caption{The values of $\eps$ for which the correct relic density is reproduced as a function of the pion mass, for different values of $\xi$ (coloured lines) and $\alpha_d$ for $N_f=3$ (left) and $N_f=4$ (right). The blue shaded area is excluded by the thermalisation requirement \cref{eq_kin}. Dotted lines violate the Bullet cluster constraint on the self-interaction. The solid lines represent the part of parameter space where the $3\rightarrow2$ interactions are the dominant freeze-out interaction. The dashed lines are when $2\rightarrow2$ annihilations via the dark photon are 
important at freeze-out.}
\label{fig:alphaDs}
\end{figure}

\Cref{fig:alphaDs} shows the numerical results for the kinetic mixing parameter as a function of the dark pion mass that reproduces the observed relic density for $N_f=3$ (left) and $N_f=4$ (right). The red, orange and purple curves correspond to different values of $\xi =2,5,10$.  In all plots the mass splitting is set to $\delta m=10^{-3}$.  The numerical results are in excellent agreement with the analytical estimate \cref{epsilon_full}.  Along the curves three different regions can be identified; i) a part where the self-interactions are larger than allowed by the Bullet cluster constraint, which is excluded (dotted); ii) a part where the $3\rightarrow2$ interactions dominate freeze-out, and annihilations are only important at later times times (solid); and iii) annihilations are the dominant freeze-out process (dashed).

The blue region in the plots is excluded by the thermalisation requirement \cref{eq_kin}. For larger $\alpha_d$ a smaller kinetic mixing parameter is required to maintain thermal equilibrium with the SM.  The results are very similar for $N_f=3$ and $N_f=4$, although the kinetic mixing (bullet cluster) constraints are slightly weaker (stronger) for $N_f=4$.

The top plots shows the result for $\alpha_d =0.01$.  For such small gauge couplings, the resonance enhancement of the WZW interaction is negligible. DM is over produced unless annihilations are important. In fact, we see that for the parameter space allowed by the Bullet cluster constraints, the annihilations are actually so large that they always dominate freeze out. Hence, in this scenario the dark pions are WIMP rather than SIMP dark matter.

In the middle plots the gauge coupling is increased $\alpha_d =0.1$, but still resonance effects on the WZW interactions are small except for large $\xi$.
Indeed, for $\xi \sim 10$ the dark pion can be SIMP DM.  Although late time annihilations reduce the relic density somewhat -- this is what generates the slope of the solid curve as a function of mixing parameter $\eps$ -- the effect is not strong enough to allow for much smaller $\xi$ than in the `standard' scenario.

Finally, the bottom plots are for sizeable couplings $\alpha_d$, and the WZW interactions are significantly enhanced for smaller $\xi = 1-5$ as well, allowing SIMP dark matter in a perturbative set-up.  The slope of the solid part of the curves shows the impact of late time annihilations as a function of kinetic mixing. The curves asymptote to a constant value for small mixing and annihilations are negligible at all times, thus providing a lower bound on the dark pion mass for a given $\xi$.

\section{Conclusion}
\label{sec:conclusion}

We have studied  a dark sector containing dark pions and a dark photon.  The dark pions are stable and can be SIMP dark matter, that is produced by freeze-out of $3 \to 2$ WZW-interactions. The dark photon mixes kinetically with the SM sector, and can maintain thermal equilibrium during freeze out.  For a fine-tuned dark photon mass $m_V \approx 2m_\pi$ the WZW are resonantly enhanced, which opens up the possibility that the observed relic density is produced in the perturbative regime $\xi =m_\pi/f_\pi \lesssim 4\pi$ of the effective chiral Lagrangian. In addition, the pion self-interactions are resonantly enhanced and become velocity dependent, which opens up the possibility that it can address the small scale structure problems of collisionless dark matter -- this scenario is dubbed resonant self-interaction dark matter RSIDM. 

We found that the RSIDM scenario is possible for dark pion masses in the range $m_\pi \sim 250 - 600$ MeV for $N_f=4$ dark quark flavours. For $N_f=3$ flavours the RSIDM scenario is (marginally) excluded for all dark pions masses considered, because of the highly efficient dark pion annihilations. In this case the value of the mixing parameter required to reproduce the relic density is too small to maintain kinetic equilibrium with the SM. For both setups more precise calculations are required to assess the viability of the model in this region. In particular, one could allow for (partial) heating of the dark bath to study the effect on the dark bath. This is left for future research.



If we give up the demand that the self-interactions have an effect on small scale structure formation, and consequently are only constraint by an upper bound from observations of the Bullet cluster, then parameter space opens up and smaller $\xi$-values become possible for sufficiently dark gauge couplings $\alpha_d \sim1$.

\section*{Acknowledgments}
The authors thank Jordy de Vries for very useful discussions. This work was funded by an NWO-klein2 grant (OCENW.KLEIN.427).

\appendix

\section{Cross sections}

We list here the definitions of the (thermally averaged) cross sections. This appendix also serves to set the notation.

\subsection{Scattering/annihilation cross section}

The cross section for scattering with two particles in both the initial and final state, labeled by $\alpha $ and $\beta$ respectively, is
\begin{align}
  \sigma_{\alpha \to \beta}
  &= \frac1{4FS_\beta}  \( \prod_{\beta=1,2} \int_{p_\beta} \) (2\pi)^4
    \delta^4(P_\alpha-P_\beta) |\bar \M_{\alpha \to \beta}|^2
             \stackrel{{\rm CM}}{=} \int \dd \Omega\frac1{(8\pi)^2 s S_\beta}
             \frac{p_{\rm out} }{ p_{\rm in}} |\bar \M_{\alpha \to \beta}|^2
    \label{sigma_scat}
\end{align}
with $\int_p = \int \dd^3 p/(2E_p(2\pi)^3)$.  We use $P^\mu$ for 4-momenta, and $p^i$ for 3-momenta.  The second expression is valid in the center of mass frame (CM), with $p_{\rm in} = |\vec p_\alpha|$ ($p_{\rm out} = |\vec p_\beta| )$ the absolute value of the three-momentum of either incoming (outgoing) particle. $S_\beta= N!$ for $N$ identical particles in the final state, to avoid overcounting in the phase space integral.  $|\bar \M|^2$ is the amplitude averaged over initial and summed over final state particles.  The flux factor can be written as $F= E_1 E_2 |v_{1}-v_{2}| = \sqrt{(p_1.p_2)^2 -m^2_{1}
  m^2_{2}}\stackrel{{\rm CM}}{=}  p_{\rm in} \sqrt{s}$ with
$s =(E_1 +E_2)^2$ the center of mass energy squared.

\subsection{Thermally averaged cross section}

The thermally averaged cross sections can be defined in terms of the scattering rates appearing in the Boltzmann equation for the DM particle \cite{Bernal:2015bla}
%
\be
\dot n +3 H n = - \sum_{\alpha,\beta} \Delta_{\alpha \beta} (\tilde \gamma_{\alpha \to \beta} - \tilde \gamma_{\beta
  \to \alpha})  = -\langle \sigma v\rangle_{\rm ann}(n^2-n_{\rm eq}^2) - \langle
    \sigma v^2 \rangle_{3 \to 2} (n^3 -n^2 n_{\rm eq}),
    \label{BE_gamma}
\ee
with $n =n_{\rm DM}$ the number density of dark matter. We have included both  DM annihilation and $3 \to 2$ interactions.  $\Delta_{\alpha \beta} = (N^\textsc{dm} _\alpha - N^\textsc{dm} _\beta) $ is the difference between the number of DM particles in the initial ($N^\textsc{dm} _\alpha $) and final state ($ N^\textsc{dm} _\beta$). The collision terms are
\be
\hat \gamma_{\alpha \to \beta}(f_\alpha) = \frac1{S_\alpha S_\beta} \(\prod_{\alpha}\int_\alpha
N_\alpha f_\alpha \) \( \prod_{\beta}\int_\beta \)
(2\pi)^4 \delta^4(P_\alpha -
P_\beta) |\bar {\cal M}_{\alpha \to \beta}|^2 
\label{gamma1}
\ee
with $f_\alpha,\, N_\alpha$ the distribution functions and degrees of freedom of the initial states, and $S_\alpha\, (S_\beta) =N!$ for $N$ identical particles in the initial (final) state.  Assuming kinematic equilibrium for the DM and chemical equilibrium for all other particles gives the relations
\be
\hat \gamma_{\alpha \to \beta} (f_i) =\(\frac{n}{n_{\rm
    eq}}\)^{N^\textsc{dm} _\alpha} \gamma_{\alpha \to \beta} (f_i^{\rm eq}), \qquad \gamma_{\alpha \to \beta} =\gamma_{ \beta \to\alpha }
\label{gamma2}
\ee
with $\gamma_{\alpha \to \beta} \equiv \hat \gamma_{\alpha \to \beta} (f^{\rm eq}_\alpha) $ and $f_\alpha^{\rm eq} = e^{-E_\alpha/T}$ the Maxwell-Boltzmann distribution. We can then express the thermally averaged cross sections appearing  in  the Boltzmann equation \cref{BE_gamma} in terms of the collision rates as follows
\be
\langle \sigma v\rangle_{\rm ann} = \frac{2 \gamma_{\rm ann}}{n_{\rm
    eq}^2},
\quad
\langle \sigma v^2\rangle_{3\to2} = \frac{\gamma_{3 \to 2}}{n_{\rm
    eq}^3}.
\label{sigma_ave}
\ee
%

For annihilations the momentum integrations in $\gamma_{\rm ann}$  can be partially done, and the final expression is given in terms of one remaining integral over the center of mass energy \cite{Gondolo:1990dk,Edsjo:1997bg}. 
The thermally averaged cross section is
\begin{align}
  \langle \sigma v\rangle_{\rm ann} 
&= \frac{1 }{2
  T m_1^2m_2^2 K_2(\frac{m_1}{T}) K_2(\frac{m_2}{T})}\int_{(m_1+m_2)^2}^\infty \dd s
 \, K_1(\frac{\sqrt{s}}{T}) (p_{\rm in} 
E_1E_2 v_{\rm m\o l}
                           \sigma)
    \label{sigmav}
\end{align}
%
with the M\o ller velocity related to the flux factor as $F=(E_1 E_2) v_{\rm m\o l} $, and the factor $\Delta_{\rm ann}/S_\alpha =1$ is set to unity.  The equilibrium number density is defined as
\be
n_\alpha^{\rm eq} = \frac{N_\alpha}{ (2\pi)^{3}}
\int \dd^3 p_\alpha f_\alpha^{\rm eq} = \frac{N_\alpha m_\alpha^2 T}{2\pi^2}K_2(\frac{m_\alpha}{T} )
\stackrel{m_\alpha\gg T}{=} N_\alpha \( \frac{m_\alpha T}{2\pi}\)^{3/2} \e^{-m_\alpha/T},
\label{n_eq}
\ee
with the last expression valid in the non-relativistic limit.
For $m_1 =m_2 \equiv m$ we can rewrite this in dimensionless variables
\begin{align}
\langle \sigma v \rangle_{\rm ann} &= \frac{4x \Delta }{S_\alpha
  K_2(x)^2}\int_{1}^\infty \dd \tilde s
 \, \sqrt{\tilde s} (\tilde s-1) K_1(2\sqrt{\tilde s}x) 
                           \sigma(\tilde s),
                           \label{sigmav_eq}       
\end{align}
with $\tilde s =s/(4m^2)$ and $x
= m/T$.

\subsection{S-channel resonance and narrow width approximation}

If the DM interactions are mediated by a massive meditor particle -- in our case, the dark photon -- there will be an s-channel resonance for momenta that the mediator is nearly on shell $s \approx m_V^2$. To incorporate this effect, we include the decay rate in the dark photon propagator
\be
D_{\mu\nu}(P^2)=\frac{-ig_{\mu\nu}}{P^2 - m_V^2+i\eps} \to \frac{-ig_{\mu\nu}}{P^2 - (m_V -i
  \frac12{\Gamma})^2} \approx \frac{-ig_{\mu\nu}}{P^2 - m_V^2 +i m_V
  {\Gamma}}
\label{prop_res}
\ee
where we used that $\Gamma^2 \ll m_V^2$.  $\Gamma =\Gamma_d+ \Gamma_v$ is the total decay width of the dark photon, which is the sum of the decay rate into pions and decay rate into SM fermions, i.e. into the dark and visible sector.  In the resonance limit the most enhanced terms in the cross section will be $\propto D_{\mu\nu}(s)^2$, which can be evaluated in the narrow width approximation
\be
\frac{1}{(s-m_V)^2+m_V^2 \Gamma^2} \approx \frac{\pi}{m_V \Gamma} \delta(s-m_V^2) + \O(\Gamma^2/m_V^2).
\label{NWA}
\ee
%

\section{Pion self-interactions}
\label{A:self_interactions}

In this appendix we calculate the pion self-interaction cross section $\sigma_{\rm SI} =\sigma({\pi\pi \to \pi\pi})$, which has contributions from 4pnt self-interactions and from dark photon exchange.

\subsection{Amplitude}


\paragraph{Dark photon mediated self-interaction}
The 4pnt pion interaction follows from \cref{L_pi}
\be
\L \supset  \frac{1}{3f_\pi^2}\(2\pi^a \partial \pi^b
\pi^c \partial \pi^d -2 \pi^a  \pi^b \partial 
\pi^c\partial\pi^d+m_\pi^2  \pi^a \pi^b\pi^c \pi^d\) \(
\Tr[T^a T^b T^c T^d] + {\rm perm.}\).
\ee
There are $4!$ possible orderings of the pions $a,b,c.d$. Consider first the amplitude for the $\{acbd\}$-term plus the cyclic permutations:
\begin{align}
  \M^{4{\rm pnt}}_{\{acbd\}}&= -\frac{4\Tr[T^a T^c T^b T^d]}{3f_\pi^2} \big(
                         (P_c \cdot P_d +P_a \cdot P_b)
                         +\frac12 (P_b \cdot P_d+P_c \cdot P_b +P_a
                         \cdot P_c+P_d \cdot P_a) - m_\pi^2\big) \nn \\
  &= -\frac{2}{f_\pi^2}\Tr[T^a T^c T^b T^d] \(
  s-2m_\pi^2\) ,
\end{align}
where we took $P_a, P_b$ as incoming momenta, and $P_c, P_d$ as outgoing ($\partial \pi^a \to -i P_a$, $\partial \pi^c \to i P_c$), and on the 2nd line we used the Mandelstam variables
\begin{align}
s = (P_a+P_b)^2 ,\quad 
t= (P_a-P_c)^2 ,\quad 
u= (P_a-P_d)^2. 
\label{mandelstam}
\end{align}
The results are similar for the other possible permutations, and the total amplitude is  \cite{Cline:2013zca}
\begin{align}
  \M_{abcd}^{4{\rm pnt}}={\cal M}({\pi^a \pi^b \to \pi^c \pi^d})
  = &-\frac{2}{f_\pi^2}
    \( \Tr[T^a T^b T^c T^d]+(b \leftrightarrow d) \) \(
    t-2m_\pi^2\) \nn \\ &-
    \frac{2}{f_\pi^2}\( \Tr[T^a T^c T^b T^d]+(c \leftrightarrow d) \) \(
                          s-2m_\pi^2\)
      \nn \\ & -
    \frac{2}{f_\pi^2}\( \Tr[T^a T^c T^d T^b]+(b \leftrightarrow c) \) \(
               u-2m_\pi^2\)
               \label{Mabcd_1}
\end{align}

\paragraph{Dark photon mediated self-interaction}
The pion-dark photon interactions that follow from the covariant derivatives in the chiral Lagangian \cref{L_pi} can be written in the form
\begin{align}
\L \supset
&=-2ig_dV_\mu   \(\pi^a (\partial \pi^b) -(\partial\pi^a) \pi^b\)
            \Tr\([T^a ,T^b] Q] \) .
\end{align}
The $(V2\pi)$-vertex interaction and dark photon propagator (in Lorentz gauge) are then
\be
\A_{ab}^\mu= 2ig_d(P_a-P_b)^\mu \Tr([T^a,T^b]Q), \qquad
D_{\mu\nu} (P^2)= \frac{-i g_{\mu\nu}}{P^2-m_{V}^2+i\eps}
\label{Apipi}
  \ee
  with both momenta $P_a, P_b$ incoming.
The amplitude for $\pi\pi \to \pi\pi$ scattering via dark photon exchange is
\begin{align}
{\cal M}_{abcd}^{V} 
              &=      i\A_{ab}^\mu D_{\mu\nu} (s) \A_{cd}^\nu
             -       i\A_{ac}^\mu D_{\mu\nu} (t) \A_{bd}^\nu
              -       i\A_{ad}^\mu D_{\mu\nu} (u) \A_{bc}^\nu \nn \\
                    &= 4g_d^2
                       \[\frac{(t-u)}{s-m_{V}^2+i\eps}C_{abcd} +
              \frac{(s-u)}{t-m_{V}^2+i\eps}C_{acbd}+
                      \frac{(s-t)}{u-m_{V}^2+i\eps}C_{adbc}\]
                          \label{Mabcd_2}
\end{align}
with color factor
\be
C_{abcd} \equiv \Tr([T^a,T^b]Q) \Tr([T^c,T^d]Q).
\label{Cabcd}
\ee
Resonant scattering arises for $P^2 \approx m_V^2$, and we include the decay width in the propagator \cref{prop_res} to describe this.

\subsection{Cross section}

The matrix element squared summed over final and averaged over initial states is $|\bar {\cal M}|_{\rm SI}^2 = \frac1{N_\pi^2} \sum_{abcd} |{\cal M}_{abcd}|^2 $ with $N_\pi = (N_f^2-1)$ the number of pions. The amplitude is the sum of the 4pnt interaction \cref{Mabcd_1} and the photon exchange contribution \cref{Mabcd_2}. The latter is subdominant, except for momenta near the $s$-channel resonance $s \approx m_V^2$; we can then neglect interference terms and the non-resonant contributions, and approximate the amplitude
\begin{align}
  |\bar {\cal M}_{\rm SI}|^2 
    &\approx \frac1{N_\pi^2} \sum_{abcd}\( |{\cal M}^{4{\rm   pnt}}_{abcd}|^2+|{\cal M}^{\rm res}_{abcd}|^2\)
\end{align}
with ${\cal M}^{\rm res} = {\cal M}^{V}|_{s\approx m_V^2}$ the s-channel resonance contribution from the photon exchange diagram.

\paragraph{4pnt self-interaction}
Starting with the 4-pnt contribution, the amplitude squared can be calculated using the \texttt{Feyncalc} Mathematica program \cite{Shtabovenko_2020,Shtabovenko_2016,Mertig:1990an}
\begin{align}
  |\bar {\cal M}_{\rm SI}
  ^{4{\rm pnt}}|^2 
  = 6 \kappa_{\rm SI} \frac{m_\pi^4}{f_\pi^4} -
 \kappa_{{\rm SI},2}\frac{ (s t+t u+u
  s)}{f_\pi^4},  
\end{align}
with
\be
\kappa_{\rm SI} 
=\frac{ (3 N_f^4-2 N_f^2+6)}{3N_f^2 (N_f^2-1)}  =1 +\O(N_f^{-1}),\quad
\kappa_{{\rm SI},2} 
=\frac{N_f^2}{(N_f^2-1)} =1 +\O(N_f^{-1}).
\ee
%
The cross section \cref{sigma_scat} for pion scattering is 
\begin{align}
  \sigma_{\rm SI}^{\rm 4pnt}
         = \frac{m_\pi^4}{\pi S_f f_\pi^4 s}  \( \frac{6 \kappa_{\rm SI}  }{16} +{ \kappa_{{\rm SI},2} (m_\pi^2
                p^2+\frac56 p^4) }\) \approx
                \frac{3\kappa_{\rm SI}  \xi^4}{64 \pi m_\pi^2 }
                \label{sigma_4pnt}
\end{align}
with $S_f =2$ for two identical particles in the final state, and $p=|\vec p|$ the incoming momentum of either pion in the CM frame. The last expression is valid in the non-relativistic limit $p^2 \ll m_\pi^2$ in which the velocity-independent s-wave contribution dominates, and $s \approx 4m_\pi^2$.

The result agrees with \cite{Choi:2018iit}, but differs a factor of 8 with \cite{Hochberg:2014kqa} (after rescaling $f_\pi^{\rm them} =2 f_\pi$). For $N_f=2$ it reproduces the results of \cite{Cline:2013zca} (except for the sign in front of the $\frac56 p^4$ term) but not for other $N_f$.


\paragraph{Dark photon mediated self-interaction}

The diagram for photon exchange is negligble except near the $s$-channel resonance, where the amplitude can be approximated by the s-channel diagrams, and 
\be
 |\bar {\cal M}_{\rm SI}^{\rm res}| ^2 
  =\frac{256 C_4^2 g_d^4 }{N_\pi^2}\frac{ p^4 \cos^2 \theta}{(s-m_{V}^2)^2 +m_{V}^2
    {\Gamma}^2}  \qquad \Rightarrow \qquad
  \sigma^{\rm res}_{\rm SI} 
  = \frac{16C_4^2 g_d^4 }{3\pi s  S_fN_\pi^2}\frac{p^4}{(s-m_{V}^2)^2 +m_{V}^2
    {\Gamma(p)}^2}
\ee
with $S_f=2$ amd $\Gamma(p)$ the running photon decay width, computed in \cref{A:decay}.  The color factor \cref{Cabcd} summed over flavors is $C_4^2 = \sum |C_{abcd}|^2 = 4^2 \, (8^2)$ for $N_f=3 \, (4)$.
%

The cross section can be rewritten in the more familiar Breit-Wigner form.  To do so, expand the momentum as $ p = \frac12 m_\pi v = \frac14 m_{V} v+{\cal O}(\delta m)$ with $v$ the relative velocity. The resonance velocity follows from \cref{p_res}, which gives $v_R \approx 2 \sqrt{\delta m}$ for dark photon masses \cref{mV}.
We further define the velocity dependent and resonant kinetic energies \cite{Chu:2018fzy}
\be
E(v) = \frac{ p^2}{m_\pi} =\frac14 m_\pi v^2, \qquad
E(v_R) = m_{V} -2 m_\pi =m_\pi \delta m.
\label{Ev}
\ee
We can then rewrite the resonant cross section in Breit-Wigner form 
\begin{align}
  \sigma^{\rm res}_{\rm SI} &=
           \frac{4\pi S}{m_\pi E(v)}
           \frac{ \Gamma_d(v)^2/4}{  (E(v)-E(v_R))^2 + {\Gamma(v)^2}/{4}}
                  , \qquad S = \frac{3S_f}{N_\pi^2}
                  \label{self_res}
\end{align}
where we used the non-relativistic approximation $s = m_{V}^2 + {\cal O}(v^2)$. The numerator is written in terms of the decay width into dark pions $\Gamma_d$ using the explicit expression \cref{Gamma_d}.

Ref. \cite{Chu:2018fzy} list a numerical factor $S =N_{V}/N_\pi^2$ for the ratio of mulitplicities of the resonance dark photon and DM particles; we find here an additional $S_f =2$ symmetry factor.

\subsection{Thermally averaged cross section}

The thermally averaged self-interaction cross section is \cite{Chu:2018fzy}
\be
\langle\sigma v\rangle _{\rm SI} = \sigma _{\rm SI} ^{4{\rm pnt}} \langle v\rangle +\langle\sigma _{\rm SI} ^{\rm res} v\rangle =  \sigma^{4{\rm pnt}} \langle v\rangle +
\int_0^{v_{\rm max}} f(v,v_0) (\sigma _{\rm SI} ^{\rm res} v) \dd v, 
\ee
with  the velocity distribution $f(v,v_0) = (4v^2)/(\sqrt{\pi} v_0^3)\e^{-v^2/v_0^2}$ 
taken as a Maxwell-Boltzmann distribution truncated at the halo
escape velocity $ v_{\rm max}$, and $v_0$ a constant implicitly defined via
$\langle v\rangle \simeq 2v_0/\sqrt{\pi}$.  Here we used that the self-interaction cross section is the sum of the (dominantly) s-wave 4pnt interaction and s-channel resonance contribution  \cref{sigma_4pnt,self_res}.
The resonance contribution can be calculated in the narrow width approximation 
\cref{Ev}:
%
\be
\langle\sigma v\rangle_{\rm SI}^{\rm res} =  \frac{64\pi^{3/2} S\,
\Gamma_d(v_R)  B_d(v_R) }{m^3_\pi } \frac{\e^{-v_R^2/v_0^2}}{v_0^3}.
\label{self_thermal}
\ee
with $B_d = \Gamma_d/\Gamma$ the branching ratio for decay into dark sector pions.
The decay rate into dark pions (p-wave) \cref{Gamma_d} and SM fermions (s-wave) \cref{Gamma_SM} can be parameterized as $\Gamma_i (v)= m_V \gamma_i v^{n_i}$ with $i=d,v$ for decay into dark and visible sectors, and $m_V \approx 2m_\pi$.  This factors out the explicit velocity dependence of the $(\Gamma_d B_d)$-factor in the thermally averaged cross section.



\section{Dark photon decay rate}
\label{A:decay}

The dark photon can decay into dark pions, and in SM fermions and pions via kinetic mixing with the SM photon.

\subsection{Dark photon decay rate into dark pions $\Gamma_{{V} \to \pi \pi}$.}

The dark photon decay rate in dark sector particles is $\Gamma_d =\Gamma({{V} \to \pi \pi})$. The decay into dark pions is mediated by the vertex interaction \cref{Apipi}, with amplitude
\be
{\cal M}_{ab}^\mu 
= -2 g_d(P_a-P_b)^\mu C_{ab}, \qquad C_{ab}=\Tr([T^a,T^b]Q)
\ee
with $P_a,P_b$ the 4-momenta of the outgoing pions.
The amplitude squared  (summed over final states, and averaged
over intial photon polarizations states) in the CM frame is
\begin{align}
|{\cal \bar M}|_d^2&= -\frac13   g_{\mu\nu}  \sum_{ab} {\cal M}_{ab}^\mu {\cal
              M}_{ab }^{\nu *} 
                     =\frac{16 C_4}{3} g_d^2 p_{\rm out}^2
\end{align}              
with $p_{\rm out} =|\vec p_a|=|\vec p_b|$ the
final state 3-momentum, and
color factor $C_2^2 =\sum_{ab} |C_{ab}|^2 = C_4 $.
The decay rate becomes
\begin{align}
\Gamma_d(p_{\rm out}) 
  = \frac{ p_{\rm out}}{S_f 8\pi m_{V}^2} |{\cal
  \bar  M}|_d^2 
  = \frac{8 C_4 \alpha_d p_{\rm out}^3}{3 S_f m_{V}^2}
  \label{Gamma_d}
\end{align}
with $\alpha_d =g_d^2/(4\pi)$ and $S_f=2$ for two identical
particles in the final state.

The resonance momentum (taking the intial
 state photon at rest) is
\be
p_R^2 = \frac14 m_{V}^2\(1-\frac{4
  m_\pi^2}{m_{V}^2}\) =m_\pi^2 \delta m + \O(\delta m^2)
\label{p_res}
\ee
where we used the parameterization \cref{mV} in the last step, and assumed the dark photon mass is close to resonance $\delta m^2 \ll m_\pi^2$. 
Evaluating the decay rate at resonance $p_{\rm out}=p_R$ gives
\be
\Gamma_d(p_R) = \frac{2C_4 \alpha_d}{3 S_f}m_\pi\delta m^{3/2}.
\label{GammaR}
\ee

\subsection{Dark photon decay rate in SM particles.}

The dark photon can decay into SM electrons via kinetic mixing and $\Gamma_s =\Gamma({{V} \to \bar f f})$ with $f$ the electron; for large enough DM mass the decay into muons and charged SM pions also becomes kinematically accessible.  The decay rate into SM pions will be of the form \cref{Gamma_d} with $\alpha_d \to \eps^2 \alpha$ and instead of $C_4\to C_4^{\rm QCD} =1/2$ the appropiate color factor for QCD.  It is velocity suppressed compared to decay into fermions, which is s-wave, and we neglect it in the following.

The amplitude for dark photon decay into a SM fermion is  
\begin{align}
 \M^\mu_{v} = ( ie \eps) \bar u(P_2) 
  \gamma^\mu v(P_1)
\end{align}
with $P_1,P_2$ the 4-momenta of the outgoing fermions,  $\eps$ the kinetic mixing paramer appearing in the Lagrangian \cref{L_pi}, and $e$ the electric charge of the electron.
The amplitude squared (summed over final states, and averaged
over photon polarizations) in the CM frame and decay rate are
\begin{align}
|{\cal \bar M}|_v^2
                     =\frac{1}{3} (e \eps)^2 4 m_{V}^2 \qquad \Rightarrow \qquad
\Gamma_{v}
                       = \frac{\alpha \eps^2}{3S_f} m_{V}
  \label{Gamma_SM}
\end{align}
with $\alpha =e^2/(4\pi)$, symmetry factor $S_f=1$, and
$s = m_{V}^2$.

Decay into the darks sector pions dominates and $\Gamma_d\gg \Gamma_v$ or 
\be
\label{eq:Bd}
\eps \lesssim \sqrt{\frac{ \pi \alpha_d C_4 }{ \alpha }}\delta m^{3/4}
= 8.5 \times 10^{-6} \(\frac{C_4}{4}\) \sqrt{\frac{\alpha_d }{ \alpha }}\(\frac{\delta m}{3 \times 10^{-8}}\)^{3/4},
\ee
where we set $v \to v_R$.


\section{Dark pion annilation and scattering}
\label{A:dark_photon}


Dark pions can annihilate into and scatter off SM fermions.  These processes are important for the final relic density and for keeping the dark sector in thermal equilibrium with the SM sector.

\subsection{Dark pion annihiliation into SM fermions}

Consider the annihiliation of dark pions into Standard Model electrons $\pi^a (P_1)\pi^b(P_2) \to \bar f(K_1) f(K_2)$. For large enough dark pion mass, also the decay channel into muons ($m_\pi > 105\,$MeV) and SM charged pions  ($m_\pi > 139.6\,$MeV) opens up. 

The amplitude for annihilation in a SM Dirac fermion pair $\bar f f$ is mediated by the vertex \cref{Apipi} and the coupling to the SM fermion current via kinetic mixing.
\begin{align}
i \M_{\pi\pi \to \bar f f} = -i 2 g_D (P_1-P_2)_\mu \Tr ([T^a,T^b]Q)
  \frac{-i}{s-m_{V}^2 +i\eps} (i e \eps) \sum_f \bar u(K_1) q_f
  \gamma_\mu v(K_2)
\end{align}
Averaging over initial
states and summing over the spins of the final state fermions gives
\begin{align}
  |\M|^2_{\pi\pi\to \bar f f} & = \frac{(2g_D e \eps)^2 C_4}{N_\pi^2(s-m_{V}^2 )^2} 
                                \sum_{\rm
                              spin}\(\bar u(K_1) (\slashed{P}_1
                              -\slashed{P}_2) v(K_2) \bar v(K_2) (\slashed{P}_1
                          -\slashed{P}_2) u (K_1)\) \nn \\
  &=\frac{(2g_D e \eps)^2 C_4}{N_\pi^2(s-m_{V}^2 )^2} 32 p^2\[ k^2 (1-\cos^2\theta)+ m_f^2\]
\end{align}
with color factor $C_4 =C_2^2\equiv \sum_{ab}|\Tr ([T^a,T^b]Q|^2$.
Here we denoted the CM 3-momentum of the incoming pions with $p$ and that of the outgoing fermions with $k$, and $\theta$  the scattering angle.
%
The cross section \cref{sigma_scat} becomes
\begin{align}
  \sigma_{\pi\pi\to \bar f f} & 
  = 4A_{\rm ann}
  \frac{\sqrt{s-4m_\pi^2}\sqrt{s-4m_f^2} (s+2m_f^2)}{s
  (s-m_{V}^2)^2}
                              \stackrel{m_f \to 0}{=} 
           \frac{A_{\rm ann}}{m_\pi^2}
  \frac{\sqrt{\tilde s-1}\sqrt{\tilde s} }{
                               (\tilde s-{m_{V}^2}/{(4m_\pi^2)})^2}
                               \label{sigma_ann_ff}
\end{align}
with $A_{\rm ann} = 4\pi C_4 \eps^2 \alpha_D \alpha/(3N_\pi^2)$.  In the last line we neglected the SM fermion mass, and rewrote the cross section in terms of the dimensionless variable $\tilde s = s/(4m_\pi^2)$.
%

The amplitude for annihilation in a pair of SM pions $\pi^\pm_{\rm SM} $ with is 
\begin{align}
  i \M_{\pi^a\pi^b \to \pi^c_{\rm SM} \pi^d_{\rm SM}   } = i 2 g_D (P_1-P_2)_\mu \Tr ([T^c,T^c]Q)
  \frac{1}{s-m_{V}^2 +i\eps} 2( e \eps)  (K_1-K_2)^\mu \Tr ([T^a,T^b]Q_{\rm QCD})
\end{align}
Averaging over intitial states and summing over the spins of the final state pions --  $c,d$ running over the generators corresponding to $\pi_{\rm SM}^\pm$ --
gives
\begin{align}
  |\M|^2_{\pi\pi\to 2\pi_{\rm SM} } 
  &=\frac{(4g_D e \eps)^2 C_4 C_4^{\rm QCD}}{N_\pi^2(s-m_{V}^2 )^2} 16 k^2 p^2 \cos^2\theta
\end{align}
with color factor $C_4^{\rm QCD} =1/2$.
The cross section can then be expressed as 
\begin{align}
  \sigma_{\pi\pi\to 2\pi_{\rm SM}} & =\frac{C_4^{\rm QCD}}{2} \(1- \frac{4m_{\pi_{\rm SM}^2}}{s^{3/2}}\)^{3/2} \sigma_{\pi\pi\to e\bar e}
\end{align}
where we neglected the electron mass.

Substituting \cref{sigma_ann_ff} in the thermally averaged cross section \cref{sigmav_eq} for annihilation into SM electrons is
\begin{align}
\langle \sigma v \rangle_{\pi\pi\to \bar e e}  
  &= \frac{\Delta_{\rm ann} 4x A_{\rm ann}}{S_\alpha m_\pi^2K_2(x)^2 } \int_1^\infty \dd \tilde s \frac{ \tilde s (\tilde s-1)^{3/2} K_1(2x\sqrt{\tilde s})}{(\tilde s-m_{V}^2/(4m_\pi^2))^2}
    \label{tildes_pion}
\end{align}
with $\Delta_{\rm ann}/S_\alpha =1$. We are interested in the thermally averaged cross section during freeze-out, in the limit $x =m_\pi/T \gg 1$. This will be dominated by the resonance contribution which we can calculate in the narrow width approximation \cref{NWA}. Including the decay width in the propagator \cref{prop_res} and defining $\tilde \Gamma = \Gamma/(2m_\pi), \, \tilde m_{V} =m_{V} /(2m_\pi)$ the thermally averaged cross section \cref{tildes_pion} becomes
\begin{align}
\langle \sigma v \rangle_{\pi\pi\to \bar e e} ^{\rm res}
  &    \approx  \frac{4x A_{\rm ann}}{m_\pi^2K_2(x)^2 } \int_1^\infty \dd \tilde s \tilde s (\tilde s-1)^{3/2} K_1(2x\sqrt{\tilde s})  \frac{\pi}{\tilde m_{V} \tilde \Gamma} \delta(\tilde s -\tilde m_{V}^2) \nn \\
  &\stackrel{x \gg 1}{\approx}\frac{8 \sqrt{\pi} x^{3/2} A_{\rm ann}}{m_\pi \Gamma } \delta m^{3/2} \e^{-\delta m\, x} 
\end{align}
where in the last step we used that $K_2(x) = K_1(x) = \sqrt{\pi/(2x)} \e^{-x}+...$ for $x \gg 1$, and expanded $ (\tilde m_{V}^2-1) \approx \delta m$.

Now expand the dark photon mass in small $\delta m$ defined in \cref{mV}, and use the explicit expression for the decay width \cref{GammaR} to get
\begin{align}
\langle \sigma v \rangle_{\pi\pi\to \bar e e} ^{\rm res}
    =\frac{32 \pi^{3/2} \eps^2 \alpha x^{3/2}B_d}{N_\pi^2 m_\pi^2}\e^{-\delta m \, x} +\O(\delta m)
    \label{sigma_ann_res}
 \end{align}   
 with $B_d = \Gamma_d/\Gamma$ the branching ratio for decay into the dark sector.
 We can then write the full thermally averaged annihilation cross section as
 \be
\langle \sigma v \rangle_{\rm ann} 
=g_{\rm ann}\frac{32 \pi^{3/2}\eps^2 \alpha x^{3/2}B_d}{N_\pi^2 m_\pi^2}\e^{-\delta m \, x} +\O(\delta m)
\label{sigma_ann}
\ee
with $g_{\rm ann}$ incorporating the degrees of freedom the dark pion can annihilate in. Below the muon threshold this is only electrons and $g_{\rm ann}=1$. Including muons and SM pions 
\be
g_{\rm ann} = 1 + \Theta (m_\pi - m_\mu) \sqrt{1-\frac{m_\mu^2}{m_\pi^2}}\(1+\frac{m_\mu^2}{2m_\pi^2}\)
+ \Theta(m_\pi - m_{\pi_{\rm SM}}) \frac{C_4^{\rm QCD}}{2} \(1-\frac{m_{\pi_{\rm SM}}^2}{m_\pi^2}\)^{3/2}
\ee
with $\Theta$ the Heaviside step function.


\subsection{Pion-electron scattering}

Consider scattering of dark pions with SM particles, which for light DM is
dominated by electron scattering via the reaction
$\pi^a(P_1) f(K_1) \to \pi^a(P_2) f(K_f) $. 
\begin{align}
i \M_{\pi f\to \pi f} = -i 2 g_d (P_1+P_2)^\mu \Tr ([T^a,T^b]Q)
  \frac{-i}{t-m_{V}^2 +i\eps} (i e \eps) \bar u(K_2) 
  \gamma_\mu u(K_1)
\end{align}
Averaging over intitial and summing over final states gives
\begin{align}
  |\M|^2_{\pi f\to \pi f} & =\frac{1}{2N_\pi} \frac{(2g_D e \eps)^2C_4}{(t-m_{V}^2)^2}
                                \sum_{\rm
                              spin}\(\bar u(K_2) (\slashed{P}_1
                              +\slashed{P}_2) u (K_1)\bar u(K_2) (\slashed{P}_1
                              +\slashed{P}_2) u(K_2) \)
\end{align}
The spinor sum now becomes
\begin{align}
 \sum (...)&=4\[ 2(P_1+P_2). K_1 (P_1+P_2). K_2- (P_1+P_2)^2(K_1. K_2-m_f^2)\]
          \nn \\
  & \approx 16 m_\pi^2p^2  (1 + \cos \theta ) +\O(p^2)
\end{align}
with $p$ the CM 3-momenta of the incoming particles and $k$ of the outgoing particles, and $\theta$ the scattering angle between the ingoing and outgoing pion.
In the last step we set the fermion mass to zero and took the non-relativistic limit.
%
The non-relativistic cross section becomes \cite{Hochberg:2015vrg}
\begin{align}
  \sigma_{\pi f\to \pi f}
  =\frac{8C_4 \alpha_D \alpha \eps^2 m_\pi^2 p^2}{N_\pi  s m_V^4}
  \int \dd \Omega  \, (1+ \cos \theta)
=\eps^2 A_{\rm scat}\frac{p^2}{m_{\pi}^4}
\end{align}
with $A_{\rm scat}=2 \pi C_4 \alpha_D \alpha /N_\pi$ and $m_V\approx 2m_\pi$.

The total scattering cross section is written as
\begin{align}
  \sigma_{\rm scat}
   =g_{\rm scat}\eps^2 A_{\rm scat}\frac{p^2}{m_{\pi}^4},
  \label{sigma_pi_SM}
\end{align}
with $g_{\rm scat} = 1 + \Theta (m_\pi - m_\mu) 
+ \Theta(m_\pi - m_{\pi_{\rm SM}})C_4^{\rm QCD}$, where for
simplicity we have neglected the muon and SM pion masses.


\section{Cross section for $3 \to 2$ dark pion interactions}
\label{A:SIMP}

The $3\to2$ dark pion amplitude has a contribution from the 5pnt pion interaction and from dark photon exchange. The necessary vertices with an odd number of pions arise from the WZW Lagrangian \cref{L_pi}.

\subsection{Pion 5pnt interaction from the WZW-term}

The 5pnt pion interaction from the WZW term can be written 
as \cite{Hochberg:2014kqa}
\be
{\cal L}_{\rm WZW} \supset \frac{A_5}{f_\pi^5} \eps^{\mu\nu\rho
  \sigma} \Tr\[ \vecS \pi \partial_\mu \vecS\pi \partial_\nu \vecS\pi \partial_\rho
  \vecS\pi \partial_\sigma \vecS \pi\]
, \qquad A_5=\frac{2N_c}{15 \pi^2 }
\ee
with $N_c$ the number of colors of the dark QCD-like gauge group.
 There are $5!$ different  contributions to
the amplitude ${\cal M}_{abc \to de}$. We can group them by their
momentum dependence
\begin{align}
i{\cal M}_{abc \to de}&=  \frac{iA_5}{f_\pi^5} \eps^{\mu\nu\rho
  \sigma}
\big( P_\mu^a  P_\nu^b 
  P_\rho^c P_\sigma^d f_{e} + P_\mu^b  P_\nu^c 
  P_\rho^d P_\sigma^e f_{a}  + P_\mu^c  P_\nu^d 
  P_\rho^e P_\sigma^a f_{b}  \nn \\ & \hspace{1.4cm} + P_\mu^d  P_\nu^e 
  P_\rho^a P_\sigma^b f_{c}  + P_\mu^e P_\nu^a 
  P_\rho^b P_\sigma^c f_{d} \big)
\end{align}
with $f_a$ coefficients that each are the sum of $4!$
terms, and all momenta are taken as incoming. The color-coefficient $f_e$ is defined
as
\be
f_e = T_{e abcd}- T_{eabdc}-T_{eacbd}+T_{eacdb}+T_{eadbc}-T_{eadcb}
\label{fe}
\ee
with
\begin{align}
T_{eabcd} &\equiv\Tr\[T^eT^aT^bT^cT^d\] +{\rm
              cycl.} \; \; {\rm of} \;\; \{a,b,c,d\} \nn \\ &
= \Tr\[T^eT^aT^bT^cT^d\] -\Tr\[T^eT^bT^cT^dT^a\]
                               +\Tr\[T^eT^cT^dT^aT^b\]
                               -\Tr\[T^eT^dT^aT^bT^c\]
                               \label{A5pion}
\end{align}
That is, $f_e$ is the sum of all traces of five generators with $T_e$ fixed in the first position, and all possible permutations of the other generators (of the $\{a,b,c,d\}$ indices); the sign of each term is determined by the number of permutations away from the $\{a,b,c,d,e\}$-sequence. Likewise, all other $f_i$ coefficients can be defined.

The matrix element squared averaged over initial and summed over final states is (the calculation is done with the Mathematica package \texttt{FeynCalc}) 
\begin{align}
  |{\cal {\bar M}}_{3\to 2}|^2 &= \frac{1}{N_\pi^3} \sum_{abcde} |{\cal
  M}_{abc \to de}|^2
 =\frac{A_5^2
                 N_f(N_f^2-4)(N_f^2-1) }{N_\pi^3} \frac{F(P_i)}{f_\pi^{10}}
                 \equiv \frac{\bar A^2 F(P_i)}{f_\pi^{10}}
\end{align}  
with $N_\pi =N_f^2-1$ the number of pions and $N_f$ the number of flavors.  $F$ is a complicated momentum-dependent function, which we will evaluate for non-relativistic incoming momenta.  We parameterize the momenta in the CM frame (we set $P^d \to -P^d$ and $P^e \to -P^e$ to make them outgoing momenta with positive energy as the zeroth component of the 4-vector): 
\begin{align}
P^a &= (E_1, p_1,0,0), \nn \\
P^b &= (E_2, p_2 \cos \theta, p_2\sin \theta \cos \vp , p_2\sin \theta \sin \vp ),  \nn \\
P^c &= (E_3, -p_1-p_2 \cos \theta,- p_2\sin \theta \cos \vp
          ,- p_2\sin \theta \sin \vp ),  \nn \\
P^d &= (1/2(E_1+E_2+E_3), p_4 \cos \bar \theta,
          p_4\sin \bar \theta \cos\bar  \vp , p_4\sin \bar \theta \sin
          \bar\vp ), \nn \\
P^e &= (1/2(E_1+E_2+E_3),- p_4 \cos \bar \theta,
          -p_4\sin \bar \theta \cos\bar  \vp ,- p_4\sin \bar \theta \sin \bar\vp ), 
      \label{P_explicit}
\end{align}
The momentum $p_1$ is aligned with the z-axis.  $\Omega_2$ and $\Omega_4$ are then the solid angle of $p_2$ and $p_4$ respectively.  $p_3$ and $ p_5$ are fixed in center of mass frame, and $E_4,E_5$ are fixed by energy conservation.  The energies are $E_i=\sqrt{p_i^2+m_\pi^2}$ for $i=1,2$, $E_3^2= (p_1^2 + 2 p_1 p_2 \cos \theta +p_2^2)+m_\pi^2$, and $\sqrt{p_4^2+m_\pi^2 }= 1/2(E_1+E_2 +E_3)$. We can thus express $E_i,p_4$ in terms of $p_1,p_2$.

To get the leading order term in the limit of non-relativistic momentum for the incoming particles set $p_i = \eps p_i$ for $i=1,2$ and expand in small $\eps$. The first non-zero term arises at 4th order: $F =\eps^4 F_4(p_1,p_2,\theta,\vp,\bar \theta,\bar \vp)+ {\cal
  O}(\eps^6) $ with
\be
F_4 =\frac{3375 m_\pi^4 p_1^2 p_2^2 }{16} \sin^2(\theta) \sin^2(\bar \theta)
                                                 \sin^2(\phi-\bar \phi)
\ee
%
%
%
The transition amplitude \cref{gamma1,gamma2} then becomes
\begin{align}
\gamma_{3\to 2} 
 &= \frac{\bar A^2 }{S_\alpha S_\beta (2\pi)^{11}f_\pi^{10}}
 \int \( \frac{\dd^3 p_3 \dd^3 p_4}{\prod_i(2E_i) } \) 
p_1^2\dd p_1 p_2^2\dd p_2 N_\pi^3 f^{\rm eq}_1 f^{\rm eq}_2 f^{\rm eq}_3 
\delta(E_\alpha -
   E_\beta) \int \dd \Omega \dd \bar \Omega F_4 \nn \\
  &=\frac{125 \sqrt{5}\bar A^2  m_\pi}{1024 \pi^5 S_\alpha S_\beta f_\pi^{10}} \int \frac{\dd^3 p_3 }{(2\pi)^3 }N_\pi f^{\rm eq}_3  \int  \dd
    p_1  p_1^4 N_\pi f^{\rm eq}_1 \int \dd p_2
p_2^4 N_\pi f^{\rm eq}_2 
\end{align}
with $S_\alpha =3!$ and $S_\beta =2!$ to account for identical particles in initial and final states. The integral $\int \dd \Omega \dd \bar \Omega F_4= 750 \pi^2m_\pi^4 p_1^2 p_2^2$.  To get the final expression, we further used that in the non-relativistic limit $E_i \approx m_\pi$ for $i=1,2,3$ and $E_i \approx \frac32 m_\pi$ for $i=4,5$, which yields $\prod_i(2E_i) \approx 2^3 3^2 m_\pi^5$.  Moreover $\int \dd^3 p_4 \delta(E_\alpha - E_\beta) = \frac{3}{2\sqrt5} \int \dd^3p_4
 \delta(p_4-\sqrt5/2 m_\pi) = 3 \pi\sqrt{5}/2 m_\pi^2$. The integrations over the phase space densities give
in the non-rel limit
\begin{align}
  N_\pi \int  \frac{\dd^3 p_3 }{(2\pi)^3 }f^{\rm eq}_3   
                                                      =n_\pi^{\rm eq}, 
 \qquad
N_\pi  \int  \dd
    p_1  p_1^4f^{\rm eq}_1 
                               =
                             6 \pi^2 m_\pi T n_\pi^{\rm eq}
\end{align}
The thermally averaged cross section \cref{sigma_ave} is then
\begin{align}
  \langle \sigma v^2 \rangle^{5{\rm pnt}}_{3 \to 2}
    =\frac{\alpha_{3 \to 2}}{x^2 m_\pi^5},\quad
\alpha_{3 \to 2}=\frac{ N_c^2 \kappa_{3 \to 2}}{N_f}
      \frac{5\sqrt{5} \xi^{10}}{1536 \pi^5}, \quad
      \kappa_{3 \to 2}
      =\frac{ N_f^2(N_f^2-4) }{(N_f^2-1)^2}
      =1+\O(1/N_f)
                  \label{alpha_eff}
\end{align}
with $x=m_\pi/T$ and $\xi =m_\pi/f_\pi$. This result matches the result in 
Ref \cite{Hochberg:2014kqa} (taking into account the different definitions $f^{\rm them}_\pi= 2f_\pi$).

\subsection{Dark photon interactions from WZW term}

In the presence of a dark photon, there are additional diagrams with photon exchange contributing to the $3\to2$ cross section.  The WZW term also contains photon interactions with an odd number of pions \cref{L_pi}.  
The  $A(3\pi)$ and $(2A)\pi$ interactions are
\begin{align}
{\cal L}_{\rm WZW} 
  &\supset\frac{N_c g_d}{3\pi^2 f_\pi^3}\eps^{\mu\nu\rho\sigma} A_\mu P_\nu^a
                             P_\rho^b P_\sigma^c T_{Qabc}  -\frac{N_c g_d^2}{4\pi^2 f_\pi} \eps^{\mu\nu\rho\sigma} P_\mu^{(A_\nu)}
       P_\sigma^a A_\nu A_\rho \pi^a
T_{Qa} 
\end{align}
with $T_{Qabc} =\Tr \( Q T^aT^bT^c\)$ and $T_{Qa} =\Tr \(
Q^2  T^a\)$. Including the 5pnt pion interaction discussed in the previous subsection and the $A(2\pi)$-interaciton from the chiral Lagrangian, the relevant photon-pion couplings vertices are
\begin{align}
{\cal A}_{abcde}&\equiv i \frac{A_5}{f_\pi^5} \bar {\cal A}_{abcde}=i \frac{A_5}{f_\pi^5} \eps^{\mu\nu\rho
  \sigma}
\big( P_\mu^a  P_\nu^b 
  P_\rho^c P_\sigma^d f_{e} + P_\mu^b  P_\nu^c 
  P_\rho^d P_\sigma^e f_{a}  + P_\mu^c  P_\nu^d 
  P_\rho^e P_\sigma^a f_{b}  \nn \\ & \hspace{2.3cm} + P_\mu^d  P_\nu^e 
  P_\rho^a P_\sigma^b f_{c}  + P_\mu^e P_\nu^a 
                                      P_\rho^b P_\sigma^c f_{d} \big) \nn\\
  {\cal A}_{abc}^\mu &\equiv i \frac{A_3}{f_\pi^3}   \bar {\cal A}_{abc}^\mu = i \frac{A_3}{f_\pi^3} \eps^{\mu\nu\rho\sigma} P_\nu^a
                             P_\rho^b P_\sigma^c \(
                             T_{Qabc}+T_{Qbca}+T_{Qcab} -
                       T_{Qacb}-T_{Qcba}-T_{Qbac}\) \nn\\
{\cal A}_a^{\nu\rho} &\equiv-i \frac{A_1}{f_\pi}  \bar  {\cal A}_a^{\nu\rho} =-i \frac{A_1}{f_\pi} \eps^{\mu\nu\rho\sigma} ( P_\mu^{(A^\nu)}-P_\mu^{(A^\rho)})
        P_\sigma^a 
                       T_{Qa} \nn \\
       {\cal A}^\mu_{ab}&\equiv iA_2   \bar {\cal A}^\mu_{ab}= i A_{2\pi} (P_a-P_b)^\mu \Tr([T^a,T^b]Q)
\end{align}
with 
\be
A_{5}=\frac{2N_c}{15 \pi^2 },\quad
A_{3}=\frac{N_c g_d}{3\pi^2},\quad
A_{ 1}=\frac{N_c g_d^2}{8\pi^2},\quad
A_{2} =2 g_d
\label{Ai}
\ee
and all momenta are taken as incoming.
For the ${\cal A}_{abc}$ vertex we used that there are $3!$ different terms, corresponding to $abc +{\rm cycl.}$. We have symmetrized the ${\cal A}_a $ vertex in the two  photon legs. $Q^2=1$ For our choice of charge matrix \cref{Q}, and thus $T_{Qa} = \Tr (Q^2 T^a) = \Tr(T^a) =0$, and the $(2A)\pi$ interaction vanishes $\A^{\mu \rho}_a =0$. 

The propagator is
\begin{align}
   \quad D_{\mu\nu} (P) =\frac{ -ig_{\mu\nu}}{P^2-m_{V}^2 + i
                       m_{V} \Gamma}                    =\frac{-ig_{\mu\nu}}{\Delta(P)}=\frac{-ig_{\mu\nu}}{(4m_\pi^2)\tilde \Delta(P)}
\end{align}            
where in the last step we introduced the dimensionless propagator $\tilde \Delta=  \Delta/(4m_\pi^2)$.

\subsubsection{Amplitude}

The photon interactions give rise to 6 additional diagrams contributing to the $3\to2$ interactions  \cite{Choi:2018iit}. For our charge matrix \cref{Q} the $\A^{\mu \rho}_a =0$ vertex vanishes, and only the first three diagrams contribute.
We will calculate the amplitude for $N_c=N_f =3$. The full amplitude is 
\begin{align}
  i{\cal M}&= {\cal A}_{abcde}
             +a_1  \frac{-i {\cal A}_{abc}^\mu  {\cal A}_{de\mu}}{\Delta_{(de)}} 
             + a_2 \P_{abc}\(\frac{-i \A^\mu_{ab} \A_{cde\mu}}{\Delta_{(ab)} }  \)
          +a_3 \P_{abc;de}\(\frac{-i \A^\mu_{ce}\A_{abd\mu}}{\Delta_{(ce)} } \)
\end{align}
with $\Delta_{(ij)}=\Delta(P_i+P_j)$ the propagator of momenta $P_i$ and $P_j$.  $\P_{abc}$ means all cyclic permutations of $(abc)$ -- hence there are three diagrams contributing to $a_2$ --, and $\P_{abc;de}$  cyclic permutations of $(abc)$ times cyclic permutations of $(de)$ -- hence there are six diagrams contributing to $a_3$. The $a_i$ are included as note keeping devices and can be set to unity at any time during the calculation.
The amplitude in terms of the $\bar {\cal A}$-vertices becomes 
\begin{align}
  {\cal M}&=\frac{A_5}{f_\pi^5} \bigg[  \bar  {\cal A}_{abcde}
                     +\frac{ f_\pi^2 A_2 A_3 }{(4m_\pi^2) A_5} \( a_1 \frac{\bar {\cal A}_{abc}^\mu \bar {\cal A}_{de \nu}}{\tilde \Delta_{(de)}}  
             + a_2 \P_{abc}\(\frac{\bar \A^\mu_{ab}\bar \A_{cde\nu}}{\tilde \Delta_{(ab)}}\)+ a_3 \P_{abc;de}\( \frac{\bar \A^\mu_{ce}\bar \A_{abd\nu}}{\tilde \Delta_{(ce)}}\) \) 
                      \bigg]
\end{align}
%
%
Diagram $a_2$ has an $s$-channel resonance for $m_V \approx 2m_\pi$, as the propagators $\tilde \Delta_{ij}$ with $i,j=a,b,c$ go nearly on shell.  Diagram $a_1$ can become resonant for $m_V \geq 3m_\pi$, and the propagator $\tilde \Delta_{de}$ can be put on shell.  This case was analysed in \cite{Choi:2017mkk}.  We will here not consider it any further, and instead focus on lighter dark photon masses.

Using the same momentum parameterization as before \cref{P_explicit}, the amplitude squared can be written as
\begin{align}
  |\bar \M|^2 &=\frac{3375 \bar A^2 m_\pi^4}{16 f_\pi^{10}}
 p_1^2 p_2^2\sin^2(\theta) \sin^2(\bar \theta)
                \sin^2(\phi-\bar \phi)  (1+ X)= |\bar \M|_{5{\rm pnt}}^2   (1+ X)
                \label{X}
\end{align}
with as before $\bar A^2 =A_5^2 N_\pi^{-2} N_f(N_f^2-4) \stackrel{N_f=3}{=} \frac{15}{64}A_5^2$, and $X$ parameterizing the photon-exhange corrections. 

\subsection{Resonance contribution from $m_V \approx 2m_\pi$}
In the limit that $m_V \approx 2m_\pi$ the propagators $\tilde \Delta_{(ij)}$ with $i,j=a,b,c$ are resonantly enhanced. The resonance contribution is dominated by the $\tilde \Delta_{(ij)}^{-2}$ terms in the amplitude squared proportional to $a_2^2$. Dropping all subdominant terms the correction \cref{X} becomes 
\begin{align}
X^{\rm res} &= \(\frac{ \pi \alpha_d}{\xi^2} \)^2 a_2^2
              \[\frac{128}{45} \sum_{I} \frac1{\tilde \Delta_{(I)}^2}-\frac{4}{27} \sum_{I\ne J} \frac1{\tilde \Delta_{(I)}\tilde \Delta_{(J)}}\]  
  \label{Xres}
\end{align}  
with $I,J=ab,bc,ca$.
Defining
\be
\F(h) \equiv \int\dd p_1\,  p_1^4N_\pi f^{\rm eq}_1 \int \dd p_2 \, p_2^4 N_\pi  f^{\rm eq}_2 \int \dd \cos \theta  \sin^2( \theta) h(p_1,p_2,\cos \theta)
\label{scriptF}
\ee
The transition amplitude can be written as 
\begin{align}
  \frac{\gamma^{\rm res}_{3\to 2}}{\gamma _{3\to 2}^{5{\rm pnt}}}  =
  \frac{ \F( X^{\rm res})}{ \F( 1)}, \qquad
  \F( 1)=\frac{6 \pi N_\pi^2 \e^{-2 x} m_\pi^{10} }{x^5}
\end{align}

Consider first the $\tilde \Delta_{(I)}^{-2}$ terms, which can be evaluated in the narrow width approximation \cref{NWA}
\be
\frac{1}{\tilde \Delta^2_{(I)}} 
= \frac{(4m_\pi^2)^2}{( s_{I} - m_\gamma^2)^2 + \ m^2_{V}  \Gamma^2} \approx
\frac{\pi (4m_\pi^2)^2}{ m_{V}  \Gamma} \delta( s_{I} -m_{V}^2)
\ee
with $I=ab,bc,ca$ and $s_{ab} =(P_a +P_b)^2$ etc.  The $x=\cos \theta$ integral in $\F(\tilde \Delta_{(I)}^{-2})$  becomes of the form
\be
\int \dd x\, (1-x^2) \delta(s_{I} -m_{V}^2) 
=\sum \frac{(1-x_0^2)}{|s_I'(x_0)|},
\label{delta_x}
\ee
where the sum is over the roots $x_0$ of $s_{I} -m_{V}^2=0  $. It will be useful to redefine the momenta
\be
(ab): \;\;  p_\pm = \frac1{\sqrt{2}}(p_1 \pm p_2); \qquad
(bc): \;\;  p_\pm = \frac1{\sqrt{2}}(p_1 \pm 2p_2); \qquad
(ac): \;\;  p_\pm = \frac1{\sqrt{2}}(2p_1 \pm p_2).
\label{pplus}
\ee
for $I = ab,bc,ac$ respectively.  Then for all $I$ we get
\begin{align}
   |x_0| = \frac{p_{+}^2 + p_{-}^2 - 4 m_\pi^2 \delta m }{p_{+}^2 - p_{-}^2} ,\quad  |s_I'(x_0)| = {p_{+}^2-p_{-}^2} ,
  \quad |p_{+}| \geq  m_\pi \sqrt{2\delta m} \geq| p_{-}|
  \label{x0}
\end{align}
up to $\O(\delta m^2)$ corrections. The constraint on the momentum range  arises from requiring $|\cos \theta| \le 1$; only small $p_-$-momenta can hit the resonance. A further suppression comes from the $(1-x_0^2) \propto \delta m$ in \cref{delta_x}, in the limit that $|p_-| \ll p_+$.
Putting it all together 
\begin{align}
\F(\tilde \Delta_I^{-2}) &= \frac{\pi (4m_\pi^2)^2}{ m_{V}  \Gamma}\int\dd p_1\,  p_1^4N_\pi f^{\rm eq}_1 \int \dd p_2 \, p_2^4 N_\pi  f^{\rm eq}_2  \frac{(1-x_0^2)}{|s_I'(x_0)|} \nn \\
&\stackrel{p_+ \gg |p_-|}{\approx}
\rho_I \frac{ 8 \pi m_\pi^6 \delta m}{ m_{V}  \Gamma}\int^\infty \dd p_+p_+^4\, \int_{-m_\pi \sqrt{2\delta m}}^{m_\pi \sqrt{2\delta m}}  \dd p_-     N_\pi^2  f^{\rm eq}_1 f^{\rm eq}_2 \nn \\
                    & = \tilde \rho_I \frac{ 48\pi \sqrt{\pi} N_\pi^2  m_\pi^{12} (\delta m)^{3/2}\e^{-2 x}}{ m_{V}  \Gamma x^{5/2}}
             = \tilde \rho_I \frac{ 36 S_f B_d \pi \sqrt{\pi} N_\pi^2  m_\pi^{10}\e^{-2 x}}{ C_4 \alpha_d x^{5/2}}       
\end{align}
On the 2nd line $\rho_I=1$ for $I=ab$ and $\rho_I =2^{-5}$ for $I=bc,ac$; the suppression of the latter terms come from the factors of $2$ in the definition of $p_\pm$ in \cref{pplus} (including a factor $1/2$ from the Jacobian). On the last line $\tilde \rho_i =1$ for $I=ab$, and $\tilde \rho_i =\rho_1 (128/25)\sqrt{2/5} \approx 0.1$ for $I=bc,ac$, with the additional factor for $I=bc,ac$ arising from the different $p_\pm$-dependence of $f_i^{\rm eq}$. The final expression uses the explicit decay width \cref{GammaR} into pions, $B_d = \Gamma_d/\Gamma$ the branching ratio, and $m_V \approx 2m_\pi$. A careful inclusion of the integration boundary replaces (to first order in $\delta m$) 
%
\be
\e^{-2x} \to \e^{-2x\sqrt{\tilde s}} = \e^{-2x-\delta m x}, 
\ee
which suppresses the interactions when the center of mass energy drops below the temperature. This is the same exponential factor as found in the thermally averaged annihilation cross section \cref{sigma_ann_res}.

The $I=bc,ac$ contributions are subdominant.  Expecting the mixed terms in \cref{Xres}, which already come with a small coefficient, likewise to be subdominant, we can approximate the resonant interaction by the $\tilde \Delta_{(ab)}^{-2}$-term. Then 
\begin{align}
  \frac{\langle\sigma v^2\rangle^{\rm res}_{3\to 2}}{\langle\sigma v^2\rangle _{3\to 2}^{5{\rm pnt}}}=
  \frac{\gamma^{\rm res}_{3\to 2}}{\gamma _{3\to 2}^{5{\rm pnt}}}
  \approx  \(\frac{ \pi \alpha_d}{\xi^2} \)^2
  \frac{128}{45} 
  \frac{ \F( \tilde \Delta_{(ab)}^{-2})}{ \F( 1)}= \frac{256S_f  \pi^2\sqrt{\pi} }{15C_4}  \frac{ \alpha_d x^{5/2}}{\xi^4}
\label{SIMP_photon}
\end{align}
where we have set $a_2=1$.




 \bibliographystyle{jhep} 
\bibliography{DM}

\end{document}